\documentclass[prx, reprint, floatfix]{revtex4-2}

\usepackage{amsmath}
\usepackage{xcolor}
\renewcommand{\a}{{\bf a}}

\newcommand{\ep}{\epsilon}

\newcommand{\A}{{\cal A}}

\newcommand{\Z}{{\bf Z}}
\newcommand{\z}{{\bf z}}

\newcommand{\outcomment}[1]{}

\newcommand{\EQ}{\begin{equation}}
\newcommand{\EE}{\end{equation}}
\newcommand{\EQA}{\begin{eqnarray}}
\newcommand{\EEA}{\end{eqnarray}}

\usepackage{graphicx}
\graphicspath{ {../figures/} }
\usepackage{hyperref}

\usepackage[normalem]{ulem}
\usepackage{soul}

\begin{document}
\title{Adaptive ratchets and the evolution of molecular complexity}

\author{Tom R\"oschinger${}^{1}$, Roberto Mor\'an-Tovar${}^{2}$, Simone Pompei${}^{3}$, and Michael L\"assig${}^{2}$}
\email{Contact author. Email mlaessig@uni-koeln.de} 
\affiliation{${}^{1}$Division of Biology and Biological Engineering, California Institute of Technology, Pasadena, CA 91125 
 \\
${}^{2}$Institut f\"ur Biologische Physik, Universit\"at zu K\"oln, Z\"ulpicherstr. 77, 50937 K\"oln, Germany
 \\
${}^{3}$IFOM Foundation, FIRC Institute for Molecular Oncology, via Adamello 16, Milan, Italy}

\date{\today}

\begin{abstract}
    Biological systems have evolved to amazingly complex states, yet we do not understand in general how evolution operates to generate increasing genetic and functional complexity. Molecular recognition sites are short genome segments or peptides that can bind a cognate recognition target of sufficient sequence similarity, thereby inducing a target-dependent function. Such sites are simple, ubiquitous modules of sequence information, cellular physiology, and evolution. Here we show that recognition sites, if coupled to a time-dependent recognition target, can rapidly evolve to complex states with larger code length and smaller coding density than sites recognizing a static target. Such processes take place in a new mode of molecular evolution called an {\em adaptive ratchet}, characterized by asymmetric selection for code extensions and compressions. The underlying evolutionary model contains selection for recognition depending on the site-target binding affinity and mutations with a separation of time scales between changes of binding function and complexity changes. We show that ratchet evolution increases the number of mutational paths available for adaptive code changes, thereby accelerating the response to moving targets and facilitating refinement and innovation of recognition functions. We apply these results to the recognition of fast-evolving antigens by the human immune system. Our analysis shows how molecular complexity can evolve as a collateral to selection for function in a dynamic environment. 
\end{abstract}

\maketitle

Darwin's principle states that evolution progresses by mutations and selection. This principle describes a dynamical pathway to organisms of high fitness, the efficacy of which is modulated by other, stochastic evolutionary forces. In microbial systems, for example, fitness can increase drastically even on time scales of laboratory evolution experiments; these processes are carried by multiple mutations and occur repeatably in parallel-evolving populations~\cite{wiser2013long}. In a similar way, increasing complexity appears to be a ubiquitous feature of biological evolution. Already the onset of evolution requires the complexity of biopolymers to grow above a minimum required for autocatalytic replication~\cite{higgs2015rna,tkachenko2018onset}. In macro-evolution, a prominent example is the increase of genome and regulatory complexity in the transition from prokaryotic to multicellular eukaryotic organisms~\cite{ptashne2002genes}. Nevertheless, the evolutionary forces behind complexity have remained controversial~\cite{lynch2007frailty}. From a classical Darwinian perspective, organisms evolve complex functions because they adapt to complex environments; that is, complexity increases by positive selection. 
For example, complex gene regulatory networks have been argued to evolve in response to complex environments in adaptive, incremental steps~\cite{jenkins2010novo}. However, producing and maintaining complex features also requires a physiological machinery. This generates a fitness cost of complexity, the relative weight of which is often unknown. Recent genomic and functional experiments show that molecular complexity can arise by neutral evolution and be maintained by selection, even if there is no discernible fitness benefit~\cite{covello1993evolution, stoltzfus1999possibility, lynch2003origins, lukevs2011neutral,finnigan2012evolution,manhart2015protein,starr2018pervasive,hochberg2020hydrophobic,pillai2020origin}. On much shorter time scales, humans and other vertebrates evolve complex antibody repertoires during their lifetime; these antibodies encode a memory of past infections and serve in the immune defense against future infections by similar antigens~\cite{eisen1964affinitymaturation, Mayer2019}. Is there a universal force that drives evolving systems to higher complexity, much as selection acts to increase fitness, or are the explanations for complex outcomes as complex as the subject matter? This question has no satisfactory answer to date. We lack a general definition of complexity and a unifying theory that predicts when complexity is expected to evolve. 

In this paper, we study the evolution of molecular complexity for recognition sites in DNA or proteins. By binding a cognate molecular target, such sites perform specific regulatory, enzymatic, or signaling tasks that feed into larger functional networks of the cell. Recognition sites are simple, ubiquitous, and often quantitatively understood units of molecular function, making them an ideal subject of our study. Most importantly, a single molecular phenotype, the binding affinity to the recognition target, characterizes the functionality of a site and is the target of natural selection. At the sequence level, recognition sites are approximately digital units of molecular information. The binding specificity of the target is often summarized in a sequence motif that lists the preferred nucleotides or amino acids at each position of the recognition site. Functional sites are biased  towards sequences matching the target, which typically generates a relative information, or entropy loss, of order 1 bit per unit of functional sequence~\cite{schneider1986information,stewart2012transcription}. The code length of a recognition site can be taken as a measure of its molecular complexity. This measure is quite variable across different classes of recognition sites. Strong transcription factor binding sites in prokaryotes have typical lengths $\sim 10$ base pairs; these sites are compact functional units close to the minimal coding length required for function. This lower bound is comparable to the algorithmic complexity of a computer program predicting recognition of the target ~\cite{kolmogorov1998OnTO, wallace1999minimumlength}. Weak binding sites, often acting together with adjacent sites, spread this information to larger code length and lower coding density. Prokaryotic RNA polymerase binding sites contain about 40 base pairs, a subset of these positions carries a highly conserved recognition code~\cite{kinney2010using, brewster2012tuning}. More complex transcription units containing low-affinity sites are also common in eukaryotes~\cite{kribelbauer2019low}. Similarly, human immune recognition can be carried by compact T cell receptors binding to antigenic peptides of code length $< 10$ amino acids; more complex immune responses can involve multiple antibody lineages interacting with their cognate antigenic sites~\cite{wilson1990structural}. Immune recognition systems have recently been discussed as targets of host-pathogen co-evolution and eco-evolutionary control~\cite{nourmohammad2016host,sachdeva2020tuning,lassig2020eco}. These examples contain the main question of this paper. When is the recognition encoded in sites close to minimal code length needed for specificity? Conversely, when do we find longer and fuzzier sites, where higher molecular complexity is associated with the same function? 

Recognition sites and their complexity are the outcome of evolution. These dynamics has three kinds of elementary steps. Point mutations of the recognition site update individual letters and may affect the coding density; code extensions and compressions also change the length of the site. Moreover, in most instances, the target of recognition evolves itself. Target changes can be independent of the recognition function, as it is often the case for proteins with multiple functions. They can also be directed to change recognition; for example, the evolution of transcription factors can update regulatory networks~\cite{lozada2006bacterial, wagner2008gene,lynch2008resurrecting, perez2009evolution, voordeckers2015regulatory, friedlander2017evolution, igler2018evolutionary}. On shorter time scales, a prominent example is escape mutations of antigens from immune recognition by their hosts. Natural selection acts on the recognition function, which depends on the target binding affinity $\Delta G$, but is blind to details of site and target sequence. This has an important consequence: at a given level of $\Delta G$, the molecular complexity of sites does not carry any direct fitness benefit. Moving targets generate time-dependent selection and, if the rates of change are mutually in tune, induce a continual adaptive evolution of their cognate sites. Co-evolving sites can maintain recognition, albeit at a lower affinity $\Delta G$ than for static targets, because the adaptation of the site sequence always lags behind the evolution of the target~\cite{held2014adaptive}. Together, these dynamics shape sequence architecture, information content, and functionality of recognition sites. 

\begin{figure}[t]
  \centering
  \includegraphics[scale=1]{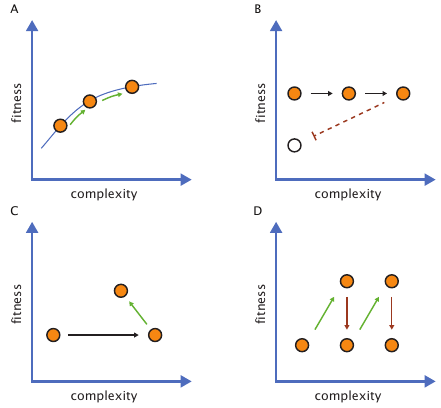}
  \caption{{\bf Evolution of molecular complexity.} The schematic distinguishes four modes of evolution: 
(A) Adaptive evolution in a static fitness landscape. Complexity-increasing changes carry a direct fitness benefit, e.g., in a complex environment. 
(B) Constraint ratchet, entrenchment. Complexity increases by neutral evolution; the complex state gets locked in by the buildup of negative selection against the reverse dynamics. 
(C) Neutral evolution inducing functional innovation. Complexity increases without a fitness benefit; complex states are stabilized by the subsequent adaptive evolution of new functions. 
(D) Adaptive ratchet. Complexity increases by adaptive evolution (green arrows) in response to a moving target; the externally driven target dynamics degrades fitness (purple arrows).
  }
  \label{fig:fig1}
\end{figure}

Here, we describe the evolution of recognition sites in a minimal, analytically solvable, yet biophysically realistic model of function and fitness. We find two evolutionary pathways towards molecular complexity: sites become long and fuzzy when the cost of complexity is low or when the target of recognition is moving. In both cases, the evolution of complex sites does not follow the classical Darwinian scenario of adaptation in a static fitness landscape (Fig.~\ref{fig:fig1}A). Rather, recognition sites evolve in a ratchet mode under time-dependent selection where opposite steps of code extension and compression do not have opposite selection coefficients. 

In the low-cost regime, the evolutionary ratchet operates by constraint: site extensions are near-neutral, selection for the recognition function builds up in the extended state and suppresses subsequent site compressions (Fig.~\ref{fig:fig1}B). Comparable evolutionary mechanisms have previously been discussed for protein and whole-genome evolution. Proteins can evolve by neutral gene duplication to more complex multimeric states that become evolutionarily stable by entrenchment: conditionally neutral processes, such as partial loss-of-function evolution in each daughter copy, make both copies dependent on each other and generate negative selection against loss of complexity~\cite{finnigan2012evolution, starr2018pervasive, hochberg2020hydrophobic, pillai2020origin}. On larger genomic scales, it has been argued that the decreased effective population sizes of multi-cellular organisms induces an initially neutral increase of sequence complexity. The more complex genome state provides a substrate for the subsequent adaptive evolution of new functions, selection on which stabilizes this state~\cite{lynch2003origins} (Fig.~\ref{fig:fig1}C). 

In the moving-target regime, we find a new mode of molecular evolution that we call an adaptive ratchet. In this mode, all extension and compression steps are driven by asymmetric, positive selection on recognition of the moving target (Fig.~\ref{fig:fig1}D). Starting from a low-complexity initial state, we show that stronger selection for extension generates a fast and robust evolution of molecular complexity that can overcome a substantial cost of complexity. Evolved sites have a specific evolutionary advantage in dynamic environments: their increased susceptibility to selection facilitates adaptation to moving targets, the refinement of existing functions, and the adoption of new functions. As an application, we discuss the benefits of complexity in the immune response to fast-evolving antigens.


\section{Model for recognition function} 

Recognition sites work in a complex molecular environment. Transcription factor binding is a well-studied case in point: functional binding at target sites competes with spurious binding to off-target in a genome-wide sequence background. This imposes stringent requirements for functionality. First, spurious recognition generated by accidental binding of target molecules to off-target sites must be sufficiently rare \cite{friedlander2016intrinsic}. 
This sets a lower bound on the information content of recognition motifs, which can be defined as the Kulback-Leibler distance between the ensemble of target-bound sites and background sequence; see Appendix~\ref{app_B}. Second, functional sites must be target-bound at physiological concentrations of the target molecule, which sets a lower bound on the target binding affinity $\Delta G$. For transcriptional regulation in prokaryotes, these functionality requirements have been shown to broadly reproduce the observed architecture of minimally complex recognition sites~\cite{gerland2002selection, berg2004adaptive,lassig2007biophysics,tuugrul2015dynamics}. They also inform the minimal model used in this study~\cite{berg1987selection}: sites have a code length $\ell \gtrsim 10$ and a sequence alphabet of $q$ equiprobable letters (with $q = 4$ for nucleotides and $q = 20$ for amino acids); 
each sequence position has one matching letter and $(q-1)$ mismatches with a reduced free energy difference $\Delta \Delta G /(k_BT) = \ep_0 \sim 1$. In this model, we can characterise recognition sites by just two summary variables: the code length $\ell$ and the number of matches, $k$, or equivalently, the coding density $\gamma = k/\ell$. 

Under conditions of thermodynamic equilibrium, recognition of the target depends on the reduced binding affinity, or free energy gap, $\Delta G /(k_B T)$. Equilibrium binding is a justified assumption whenever the target binding kinetics relaxes fast compared to the physiological time scales of target density variation \cite{morrison2021reconciling}. The recognition probability $R$ is then a standard Hill function,
\EQ
R(\gamma, \ell) = H (\Delta G - \Delta G_{50}), 
\label{R_min} 
\EE
as given in Appendix~\ref{app_A}. The functionality threshold (half-binding point) $\Delta G_{50}$ is set by the requirement of sufficient discrimination of functional sites from off-target sites in background sequence. For simplicity, we assume that background sequence can be modeled as equi-distributed random sequence, and we measure binding affinity with respect to a zero point of average random sequence. In the minimal model, the reduced binding affinity then takes the simple form $\Delta G / (k_B T) = (\gamma - \gamma_0) \ell \ep_0$, where $\gamma_0 = 1/q$ is the density of matches in random sequence. In line with empirical data, we use a threshold value $\Delta G_{50}^{\min} \sim 10 \, k_B T$ for compact binding sites under physiological conditions. In the minimal model, functional sites have at least $\ell_0 \equiv \Delta G_{50}^{\min} /(\ep_0 k_BT)$ matches more than average random sites, which requires a minimal code length $\ell_{\min} = \ell_0 / (1 - \gamma_0)$. In a heterogeneous sequence background, $\Delta G_{50}$ acquires a weak dependence on the site length $\ell$, as derived in Appendix \ref{app_A}. In the following, we use length independent approximation in the analytical calculations for simplicity, however, numerical calculations include the correction factor.

\section{Fitness model} 

In a biophysical fitness model, the recognition-dependent fitness of a site is taken to be proportional to the recognition probability, $F_r (\gamma, \ell) = f_0 R(\gamma, \ell)$. This type of fitness model generates specific fitness nonlinearities (epistasis) that can be traced in the evolutionary statistics of recognition sites~\cite{mustonen2008energy,das2020predictable}. It has been broadly used for regulatory binding sites, protein interaction sites, and protein folding~\cite{gerland2002physical,berg2004adaptive, lassig2007biophysics, zeldovich2007protein,manhart2015protein,chi2016selection,rodrigues2016biophysical,rotem2018evolution}. In this class of models, selection acts on the recognition code only through the intermediate biophysical phenotype $\Delta G$. This is an important prerequisite for the evolution of complexity: functionality of recognition imposes a lower bound on the code length, $\ell \geq \ell_{\min} \sim \Delta G_{50}^{\min}$ as discussed above, but does not generate an upper bound. 

\begin{figure}[t]
\centering
\includegraphics[scale=1]{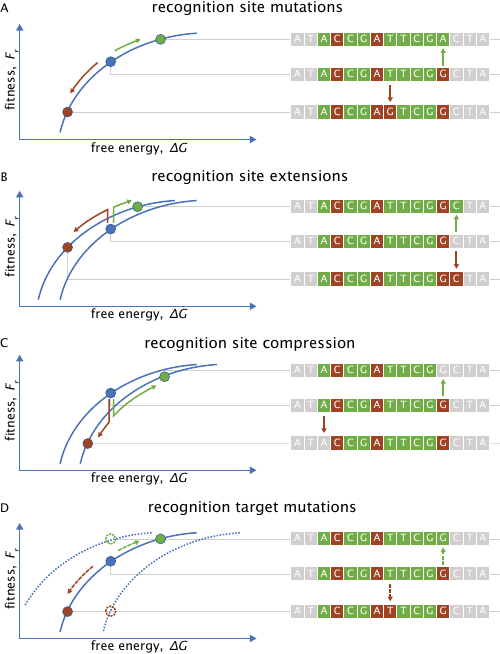}
\caption{\textbf{Evolution of molecular recognition.} Evolutionary steps change the recognition code (green squares: target matches, purple squares: mismatches). These changes affect  
the free energy of recognition, $\Delta G$, and the recognition fitness, $F_r$ (green arrows: beneficial changes, purple arrows: deleterious changes). 
 (A) Point mutations of a recognition site add or remove one letter matching the target, $k \to k \pm1$, at constant code length $\ell$. 
 (B) Site extensions and (C)~compressions add or remove one unit of length, $\ell \to \ell \pm 1$. 
 (D) Recognition target mutations make the fitness of a given recognition site explicitly time-dependent (dotted lines) and act as effective site mutations with neutral rates (dashed arrows). 
 }
 \label{fig:fig2}
\end{figure}

The (Malthusian, additive) fitness landscape $F(\gamma, \ell)$ determines the selection acting on changes of the recognition site sequence. In the minimal model, a mutation from $k$ to $(k+1)$ matches has the selection coefficient $s_\gamma = F_r (\gamma + 1/\ell, \ell) - F_r (\gamma, \ell)$. As will become explicit in the following, most functional recognition sites are found on the upper branch of the Hill function just above the half-binding point (Fig.~\ref{fig:fig2}). In this region, we can approximate the Hill function by an exponential function, which shows that selection is proportional to the recognition error, $s_\gamma = \ep_0 f_0 \, \Delta R$ with $\Delta R = 1-R$.

Other components of cell physiology, including genome replication and competing molecular functions, are expected to constrain the code length of recognition sites. Here we describe this constraint by a minimal model, choosing a linear fitness cost of genomic real estate, $F_c = - c_0 \ell$. 

Together, recognition fitness and cost of complexity define our minimal fitness model, 
\EQ
F(\gamma, \ell) = F_r (\gamma, \ell) + F_c (\ell) =  f_0 R (\gamma, \ell) - c_0 \ell.
\label{F}
\EE
This model has two system-specific parameters: the fitness effect of recognition, $f_0$, and the complexity cost parameter $c_0$, which will be used to discriminate modes of complexity evolution.

\begin{figure*}[h!]
\centering
\includegraphics[scale=1]{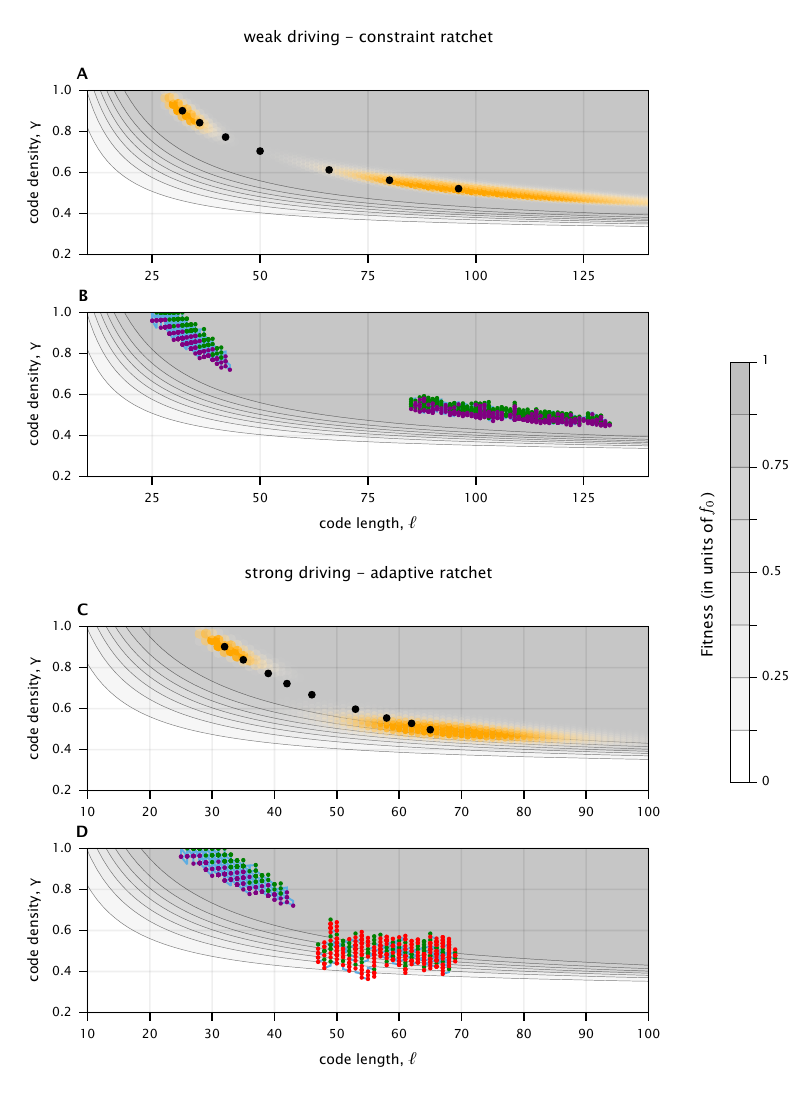}
\caption{{\bf Statistics and evolutionary trajectories of recognition sites}. 
(A, C)~Joint stationary distributions of code length and number of matches for functional sites, $Q_f (\gamma, \ell)$, for different values of the scaled complexity cost, $2N c_0$, and the driving rate, $\kappa$. Contours show the fitness landscape $F(\gamma, \ell)$, which is steepest at half-binding, $\gamma = \gamma_{50}$. 
(A)~Weak-driving regime: distributions $Q_f (\gamma, \ell)$ for $2N c_0 = 1$ (left) and $2N c_0 = 0.1$ (right) at $\kappa = 0$, maximum-likelihood points $(\gamma^*, \ell^*)(2N c_0, \kappa \! = \! 0)$ (black dots) for $2N c_0 = $ 1, 0.7, 0.5, 0.35, 0.2, 0.14, 0.1 (left to right). 
(C)~Crossover to the strong driving regime: distributions $Q_f (\gamma, \ell)$ for $\kappa = 0$ (left) and $\kappa = 15$ (right) at $2N c_0 = 1$, maximum-likelihood points $(\gamma^*, \ell^*)(2N c_0 \! = \! 1, \kappa)$ (black dots) for $\kappa = $ 0, 0.2, 0.5, 1, 2, 5, 10, 15, 30 (left to right).
(B, D)~Evolutionary paths, $(\gamma,\ell)(t)$, with marked recognition site mutations (adaptive: green, deleterious: purple) and recognition target changes (red). Sample paths are shown 
(B)~for $2N c_0 = 1$ (left) and $2N c_0 = 0.1$ (right) at $\kappa = 0$, and 
(D)~for $\kappa = 0$ (left) and $\kappa = 30$ (right) at $2N c_0 = 1$. 
Other evolutionary parameters: $\gamma_0 = 1/4$, $\ell_0 = 10$, $\ep_0 = 0.5$, $2N f_0 = 1000$, $\nu / \mu = 0.1$, $N=10^5$.\\
}
\label{fig:fig3}
\end{figure*}

\section{Evolutionary dynamics} 
We study the evolution of recognition sites in the low-mutation regime, where site sequences are updated sequentially by substitutions in an evolving population. The rates of these substitutions depend on the corresponding mutation rates and on selection coefficients scaled by an effective population size $N$, as given by Haldane's formula~\cite{haldane1927mathematical} (Appendix~\ref{app_A}). Our minimal model contains three types of updating steps. First, point mutations of the site sequence occur at a homogeneous rate $\mu$ per unit of length (Fig.~\ref{fig:fig2}A). Hence, a recognition site with $k$ matches and $\ell - k$ mismatches has a total rate $\mu_+ (\gamma, \ell) = \mu \ell \gamma \gamma_0 / (1 - \gamma_0)$  of beneficial mutations and a total rate $\mu_- (\gamma, \ell) = \mu \ell (1 - \gamma)$ of deleterious mutations (Appendix~\ref{app_A}). These changes have selection coefficients $\pm s_\gamma (\gamma, \ell)$ and substitution rates $u_\pm (\gamma, \ell)$, which depend on the fitness landscape $F(\gamma, \ell)$ introduced above.

Second, the recognition target sequence can change by extension and compression steps, which include one unit of sequence into the recognition site or exclude one unit from recognition(Fig.~\ref{fig:fig2}BC). These changes, which affect the architecture of recognition, are assumed to occur at a much lower rate than point mutations, $\nu \ll \mu$, and to be generated by external factors independent of the recognition function. Both of these features will emerge as essential for the evolutionary dynamics of complexity emerging in this system. 
Extensions add a random letter of background sequence to the site, whereas compressions exclude a random letter of the existing site sequence. In the minimal model, beneficial and deleterious extensions occur with total rates $\nu_{++} (\gamma, \ell) = \nu \gamma_0$ and $\nu_{+-} (\gamma, \ell) = \nu (1 - \gamma_0)$, respectively; beneficial and deleterious compressions occur with total rates $\nu_{-+} (\gamma, \ell) = \nu (1 - \gamma)$ and $\nu_{--} (\gamma, \ell) = \nu \gamma$. Furthermore, we assume extensions and compressions only occur at the flanks of the recognition target. We note that the evolutionary dynamics of any realistic receptor-target interface is subject to multiple steric constraints. Such constraints are likely to be strongest for internal positions of the recognition site, which is partly accounted for by restricting changes to the flanks of the recognition target. The compression and extension rates, together with the corresponding selection coefficients $s_{\pm \pm} (\gamma, \ell)$ given by the fitness landscape $F(\gamma, \ell)$, determine the total substitution rates for extensions and compressions, $v_+ (\gamma, \ell)$ and $v_- (\gamma, \ell)$ (Appendix~\ref{app_A}). For simplicity, the neutral rates of these processes are taken to be independent of the length of the sequence and there is no preference between compression and extension, i.e., they both neutral rates scale equally with $\nu$. We recall that the fitness cost $F_c$ introduced above generates an asymmetry of the corresponding substitution rates; any asymmetry of the neutral rates can be absorbed into an effective cost parameter $c_0$. 

Third, the recognition target sequence changes at a rate $\rho = \kappa \mu$ per unit of length (Fig.~\ref{fig:fig2}D). This rate is assumed to be comparable to the point mutation rate $\mu$ and to be generated by external factors independent of the recognition function similar to compression and extension mutations. The selective effects of target changes take the form of a fitness seascape, generating an explicitly time-dependent fitness of a given recognition site sequence~\cite{mustonen2010fitness}. On the recognition-fitness map $F_r(\gamma, \ell)$ and in the low mutation regime, target changes act as additional effective mutations that change the free energy $\Delta G$ and the recognition $R$ with respect to the moving target. Examples for recognition target sequence mutations are mutations in DNA binding domains of transcription factors. Experiments have shown that the effects of mutations in the DNA binding domain of LacI in \textit{E. coli} can be reduced to a change in binding energy, similar to mutations in the DNA itself~\cite{chure2019predictive}. Importantly, the selective effects of target mutations are again given by the fitness landscape $F(\gamma, \ell)$ but their rates are not: beneficial and deleterious changes occur at neutral rates $\rho_+ (\gamma, \ell) = \rho \ell \gamma \gamma_0 / (1 - \gamma_0)$ and $\rho_- (\gamma, \ell) = \rho \ell (1 - \gamma)$, respectively, generating a net degradation of recognition and driving the adaptive evolution of recognition sites. 

Clearly, the minimal evolutionary model is a broad approximation to the evolutionary dynamics of any specific receptor-target interface. The model neglects many details of actual molecular evolution processes that are not important for conclusions of this paper. Three model features turn out to be crucial for what follows. First, the sequence mutation dynamics does not introduce any bias towards higher complexity. Second, there is a separation of mutational time scales: site complexity changes take place at a much lower rate than recognition site and target mutations ($\nu \ll \mu$). Third, because selection on the recognition function depends only the binding affinity phenotype $\Delta G$, the model does not introduce any explicit fitness benefit of sequence complexity. Nevertheless, complexity can emerge as a collateral of selection for function in a non-equilibrium dynamical pathway -- an evolutionary ratchet. We will first characterize the complexity of stationary states in different parameter regimes of the evolutionary model, then derive specific dynamical pathways towards these states.

\section{Stationary states of recognition evolution} 
In an evolutionary steady state, recognition sites keep catching up with a moving target and maintain time-independent ensemble averages of binding affinity and recognition probability. In the minimal model, stationary states are characterized by time-independent averages of coding length and density, $(\gamma, \ell)$, while site and target sequences undergo continuous turnover. In general, non-equilibrium processes in a nonlinear fitness landscape are quite complicated. Remarkably, however, the steady state of the minimal recognition model can be computed analytically for arbitrary target evolution rates $\rho = \kappa \mu$. We write the steady-state distribution of coding density and code length in an ensemble of parallel-evolving populations as the exponential of an evolutionary potential, $Q(\gamma, \ell)= \exp[\Psi (\gamma, \ell)]$. Because recognition site mutations occur at a much higher rate than extensions or compressions, the coding density relaxes to a stationary state between any two code length changes and the potential takes the form $\Psi (\gamma, \ell) = \Psi (\gamma | \ell) + \Psi (\ell)$. The coding density component $\Psi (\gamma | \ell)$ is built from ratios of substitution rates for beneficial code changes and their reverse deleterious changes at constant length, $(u_+(\gamma, \ell) + \rho_+ (\gamma, \ell))/(u_-(\gamma + 1/\ell, \ell) + \rho_-(\gamma + 1/\ell, \ell))$. Similarly, the length component $\Psi (\ell)$ can be written in terms of effective substitution rates for site extensions and their reverse compressions, $\bar v_+ (\ell) / \bar v_- (\ell + 1)$; these rates are obtained by averaging the basic extension and compression rates over the conditional stationary distribution $Q (\gamma | \ell)$ or, in close approximation, by using the maximum-likelihood (ML) coding density $\gamma^* (\ell)$ (see Appendix~\ref{app_B} and Fig.~B1, Fig.~B2). Furthermore, by weighting the distribution $Q(\gamma, \ell)$ with the recognition function $R(\gamma, \ell)$, we can define a distribution of functional codes, $Q_f (\gamma, \ell)$. The full analytic solution is given in Appendix~\ref{app_B}, Eqs.~\ref{Q} -- \ref{Q_f}. 

\begin{figure*}
\centering
\includegraphics[scale=1]{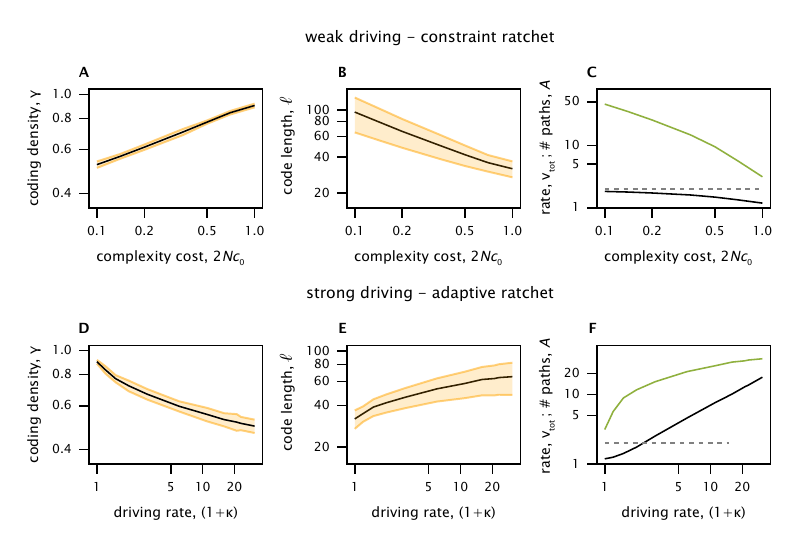}
\caption{{\bf Complexity scaling}. 
(A, D)~Coding density, ML value $\gamma^* (\kappa, 2Nc_0)$ (black), and standard deviation (orange). 
(B, E)~Code length, ML value $\ell^* (\kappa, 2Nc_0)$ (black), and standard deviation, $\Delta \ell (\kappa, 2Nc_0)$ (orange).
(C, F) ML effective number of adaptive paths, $A^* (\kappa, 2Nc_0)$ (green), and ML ratchet rate, $v^*_{\rm tot} (\kappa, 2Nc_0)$ in units of $\nu$ (black), compared to the neutral rate, $2 \nu$ (dashed). 
(A - C)~Crossover to the low cost at weak driving. All observables are plotted as functions of the scaled complexity cost $2N c_0$ at $\kappa = 0$. 
(B - F) Crossover to strong driving at unit cost. All observables are plotted as functions of the scaled driving rate $\kappa$ at unit cost, $2N c_0 = 1$. 
Evolutionary parameters as in Fig.~2. 
}
\label{fig:fig4}
\end{figure*}

Stationary distributions $Q_f (\gamma, \ell)$ for different cost parameters and driving rates are shown in Fig.~\ref{fig:fig3}; global ML points $(\gamma^*, \ell^*)$ are marked by dots. These plots display two central results of the paper. First, recognition sites for a static target ($\kappa = 0$) evolve increasing code length and decreasing coding density with decreasing cost of complexity, $2Nc_0$ (Fig.~\ref{fig:fig3}A); the cost-dependent ML values and widths of $\gamma$ and $\ell$ are shown in Fig.~4AB. Second, recognition sites for a dynamic target evolve increasing molecular complexity with increasing driving rate, $\kappa$ (Fig.~\ref{fig:fig3}C, Fig.~\ref{fig:fig4}DE).  The analytical distributions $Q(\gamma, \ell)$ (Fig.~\ref{fig:fig3}AC) are in accordance with time averages over individual trajectories obtained by simulation (Fig.~\ref{fig:fig3}BD). As a function of the target mutation rate, we observe shifts in the substitution dynamics. For static targets ($\kappa = 0$), the stationary state shows a balance of beneficial and deleterious substitutions (Fig.~\ref{fig:fig3}B; Fig.~\ref{fig:fig3}D, left), indicating evolution at equilibrium. At strong driving ($\kappa \gtrsim 1$), substitutions are predominantly adaptations to target mutations (Fig.~\ref{fig:fig3}D, right). Hence, in this regime, enhanced code complexity is associated with deviations from evolutionary equilibrium.

\subsection*{Weak- and strong-driving regimes} 
The evolutionary potential $\Psi (\gamma, \ell)$ simplifies at low and high target mutation rates, which allows for simple analytical estimates of the ML point $(\gamma^*, \ell^*)$ and gives insight in the evolutionary determinants of site complexity. We first recall that the code length $\ell$ and the conditional ML coding density, $\gamma^* (\ell)$,  set the associated selection coefficient $s_\gamma^*(\ell) \equiv s_\gamma (\gamma^* (\ell), \ell)$, on the fitness landscape $F_r (\gamma, \ell)$. At the global ML point $(\gamma^*, \ell^*)$, this relation can be written in the form 
\EQ
\ell^*  = \frac{\ell_0}{\gamma^* - \gamma_0} + O (\log (\epsilon_0 f_0 /s^*_\gamma)), 
\label{ellstar}
\EE
which says that the coding density settles on the upper branch of the recognition Hill function close to the inflection point. 

The potential $\Psi (\gamma | \ell)$ determines the ML coding density $\gamma^* (\ell)$. For weakly driven recognition sites ($\kappa \lesssim 1$), we obtain the simple form $\Psi(\gamma | \ell) = S(\gamma, \ell) + 2 N F_r (\gamma, \ell) / (1 + \kappa) - \log Z_\gamma(\ell)$ (see Appendix~B and Fig.~B1). Here, $N$ is the effective population size and $S(\gamma, \ell)$ denotes the entropy of recognition sites, defined as the log number of sequence states of given coding density and code length. In the limit $\kappa = 0$, this potential reproduces the well-known Boltzmann-Gibbs form of a mutation-selection-drift equilibrium~\cite{berg2004adaptive,sella2005application}. By maximizing $\Psi(\gamma | \kappa, \ell)$, we obtain the weak-driving balance condition for coding density, $2N s_\gamma^* (\ell)  = (\gamma^* (\ell) - \gamma_0) (1 + \kappa)/(\gamma_0(1-\gamma_0))$ (Appendix~\ref{app_B}), which says that individual code letters evolve under marginal selection, $2 N s_\gamma \sim 1$. In the equilibrium limit, this relation is accordance with general results for mutation-selection-drift-equilibria of quantitative traits~\cite{hartl1998towards,tenaillon2007quantifying,nourmohammad2013universality}. With an appropriately rescaled effective population size, it holds even if clonal interference is the dominant noise in sequence evolution~\cite{held2019survival}. Marginal selection implies that the ML coding density can get close to but cannot reach 1, which explains the substantial sequence variation observed in families of transcription factor binding sites \cite{zheng2004identification}. In contrast, under strong driving ($\kappa \gtrsim 1$), maintaining recognition requires substantial selection on individual code letters. In this regime, deleterious mutations are gradually suppressed and the potential $\Psi (\gamma | \ell)$ reflects the adaptive evolution of recognition sites in response to moving targets (Fig.~\ref{fig:fig3}D, Fig.~B1). Again by maximizing $\Psi(\gamma | \ell)$, we obtain the strong-driving balance relation for coding density, 
$2N s_\gamma^* (\ell) = (\gamma^*(\ell) - \gamma_0) \, \kappa /(\gamma_0 (1 - \gamma^* (\ell))$ (Appendix~\ref{app_B}). Importantly, the weak- and strong-driving balance conditions are highly universal: the scaled selection coefficient $2N s_\gamma^* (\ell)$ does not depend on the fitness amplitude $f_0$ or the effective population size $N$. 

Given the conditional coding density $\gamma^* (\ell)$, the potential $\Psi (\ell)$ determines the global ML point $(\gamma^*, \ell^*)$ as a function of the evolutionary parameters. In the weak driving regime, maximizing this potential yields a balance condition for code length that relates selection on recognition, $s_\gamma^*$, and the cost of complexity, $c_0$. This condition takes the form $\ell_0 (d/ d \ell) \, s_\gamma^* (\ell) |_{\ell = \ell^*} = -c_0$, which means that balance is reached once the benefit gain matches the cost of complexity. In the strong driving regime, deleterious point mutations freeze out. By balancing beneficial recognition site mutations with recognition target mutations and subsequently balancing beneficial extensions and compressions, we obtain a similar condition,  $s_\gamma^*  = (1 - (\gamma^*- \gamma_0)) \, c_0 / (\gamma_0 c_{++} - (1-\gamma^*)c_{-+})$, where $c_{++}$ and $c_{-+}$ are constants of order $0.1 - 1$ (Appendix~\ref{app_B}). This computational procedure reflects the separation of time scales between point mutations and length changes. Together with the balance condition on coding density and Eq.~[\ref{ellstar}], we obtain a closed expression for $(\gamma^*, \ell^*)$. 

In the weak-driving regime, coding density and code length scale as powers of the cost, $\ell^*/\ell_0 \sim 1/(\gamma^* - \gamma_0) \sim \sqrt{(1 + \kappa)/( 2N c_0)}$. Hence, molecular complexity can be enhanced by a reduction in the long-term effective population size, for example, in the transition from prokaryotes to multicellular eukaryotes, as suggested previously for genome evolution~\cite{lynch2003origins}. For recognition sites, this pathway to complexity builds on the universality of selection for recognition, which causes the cost-benefit ratio $c_0 / s_\gamma^*$ to decrease with a reduction in $N$. A similar universality is found  the strong driving regime: the ML point $(\gamma^*, \ell^*)$ is the solution of a quadratic equation and depends only on the ratio $\kappa/(2N c_0)$ (Appendix~\ref{app_B}). In both regimes, the analytical closure is in accordance with the full solution (Appendix~\ref{app_B}, Fig.~B2). 

\begin{figure*}[t!]
\centering
\includegraphics[scale=1]{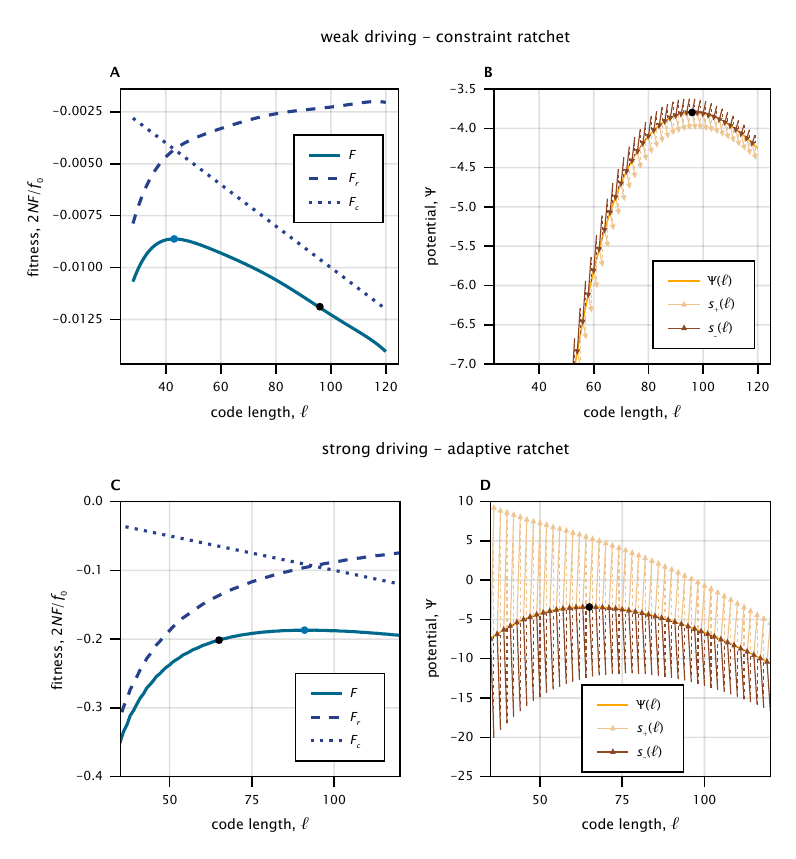}
\caption{{\bf Ratchet evolution of code length}. 
(A, C)~Effective fitness landscape for code length, $\bar F(\ell)$ (blue solid), recognition component, $\bar F_r (\ell)$ (blue dashed), and complexity cost, $F_c (\ell)$ (blue dotted). ML values $\gamma^*$ are marked by black dots; fitness values are relative to the maximum fitness $f_0$ and scaled by the effective population size $N$. 
(B, D)~Evolutionary potential for length, $\Psi (\ell)$ (orange). Arrows show ratchet selection coefficients for extension, $s_+ (\ell)$ (yellow), and for compression, $s_- (\ell)$ (brown). Dotted line segments guide the eye in joining adjacent steps. 
(A, B)~Weak-driving regime at low cost ($2Nc_0 = 0.1, \kappa = 0$); (C, D)~strong-driving regime ($2Nc_0 = 1, \kappa = 30$). See Appendix~B, Fig.~B1 for approximate potentials in both regimes. 
Evolutionary parameters as in Fig.~2. 
}
\label{fig:fig5}
\end{figure*}

\section{Ratchet evolution of code length} 
What kind of evolutionary mechanism generates the evolution of longer recognition codes? Consider first the long-term fitness effects of length changes: longer codes improve recognition but come with a cost of complexity. This tradeoff can be displayed by an effective fitness landscape for code length,  $\bar F(\ell) = \bar F_r (\ell) + F_c (\ell)$, where the recognition component $\bar F_r (\ell)$ is defined by averaging $F(\gamma, \ell)$ over the conditional stationary distribution $Q (\gamma | \ell)$ (Fig.~\ref{fig:fig5}AC, Appendix \ref{app_C}). This component shows a diminishing-return nonlinearity; its balance with a linear component $F_c (\ell)$ leads to a $\kappa$-dependent maximum of the total fitness $F(\ell)$. 

Importantly, however, code length changes do not happen alone, but they are always coupled to changes of the coding density. Moreover, given the separation of mutational time scales ($\nu \ll \mu$), the evolution of coding density settles to its stationary state before the next update of code length happens. These characteristics generate a fundamental asymmetry: code length extensions recruit new functional sequence units from random sequence, compression steps cut into existing functional sequence. Hence, extensions take place under net positive selection for recognition, compressions include a net constraint from locked-in target matches. These effects are described by average selection coefficients for extensions, $s_+ (\ell)$, and for compressions, $s_- (\ell)$ (see Appendix~\ref{app_C} and Fig.~C1). Fig.~\ref{fig:fig5}BD displays these selection coefficients as arrows on the backbone of the evolutionary potential $\Psi (\ell)$. This reveals an asymmetry that defines an evolutionary ratchet operating by selection: $s_+ (\ell)$ and $s_- (\ell + 1)$ do not sum up to 0, and both selection coefficients differ from the increment of the potential. This establishes another main result of the paper: complexity evolves in a  ratchet that clicks by different rules for forward and backward moves. 

The asymmetry of selection has a drastic consequence: the evolution of code length cannot be described by any fitness landscape that depends only on $\ell$. This would require $s_+ (\ell) = - s_- (\ell + 1) = F (\ell+1) - F(\ell)$; the arrows in Fig.~\ref{fig:fig5}BD would collapse onto their backbone curve. To capture the dynamical effects of ratchet evolution, we evaluate the total substitution rate for code length, $v_{\rm tot} (\ell) = \bar v_+ (\ell) + \bar v_- (\ell)$, and the net elongation rate, $v_e (\ell) = \bar v_+ (\ell) - \bar v_- (\ell)$. Depending on the complexity cost parameter $c_0$ and the driving rate $\kappa$, we find two different modes of ratchet evolution. 

\subsection*{Weak driving, constraint ratchets} 
In the weak-driving regime ($\kappa \lesssim 1$), we find $v_{\rm tot} < 2 \nu$, indicating a net constraint on code length evolution (the ML rate $v_{\rm tot}^*$ is shown in Fig.~\ref{fig:fig4}C, the length dependence in Appendix~C, Fig.~C1). At small $\ell$, length extensions are approximately neutral and compressions are under constraint in functional sequence units (Fig.~\ref{fig:fig5}B), generating a net elongation rate $v_e (\ell) \lesssim  \nu$ (Fig.~C1). Hence, in this regime, an initially compact recognition site ratchets up slowly to a cost-dependent ML code length $\ell^*$. This length overshoots the point of maximal fitness defined by the landscape $\bar F (\ell)$, showing again that selection is not given by gradients of this landscape (Fig.~\ref{fig:fig5}A, Appendix~\ref{app_C}). Importantly, constraint ratchets produce complex recognition sites only at very low complexity cost, $2N c_0 \ll 1$ (Fig.~\ref{fig:fig4}AB). On larger scales of functional units, comparable ratchet mechanisms have been invoked to explain the complexity of RNA editing~\cite{covello1993evolution}, cellular machines~\cite{stoltzfus1999possibility, lukevs2011neutral}, and protein domains~\cite{finnigan2012evolution,starr2018pervasive,hochberg2020hydrophobic,pillai2020origin}.

\subsection*{Strong driving, adaptive ratchets} 
In the strong-driving regime ($\kappa \gtrsim 1$), all length changes become adaptive (Fig.~\ref{fig:fig5}D), which generates fast ratchets with total click rates $v_{\rm tot} (\ell)$ and small-$\ell$ elongation rates $v_{e} (\ell)$ well above their neutral values (Fig.~\ref{fig:fig4}F, Fig.~C1). In the analytical approximation, we find a universal ML click rate 
\EQ
\frac{v^*_{\rm tot}}{2 \nu} = 2N C s_\gamma^* =  \frac{C(\gamma^* - \gamma_0)}{\gamma_0 (1 - \gamma^*)} \, \kappa , 
\label{mu_eff} 
\EE
where $C$ is a constant of order 1 (Appendix~\ref{app_C}). This shows that adaptive ratchets are driven by recognition target mutations. In contrast to constraint ratchets, adaptive ratchets generate molecular complexity also in the face of substantial costs ($2 N c_0 \sim 1$). The ratchet dynamics of recognition sites has important consequences for the evolution of recognition functions, to which we now turn.

\subsection*{Complexity increases the speed of adaptation} 
The most prominent effect of a longer code is an increased susceptibility to selection. We measure this effect by an effective number of mutational paths for adaptive evolution of recognition starting from a given site genotype, where each path is weighted by its contribution to the speed of adaptation. As detailed in Appendix~C, this number equals a suitably normalized expectation value of the population variance of the recognition trait, $A = \langle {\rm Var} (\Delta G / k_B T) \rangle / (2 \mu N \ep_0^2)$. Sites with larger $A$ adapt their code faster at a given fitness gradient $\partial F / \partial (\Delta  G)$; equivalently, such sites require less selection for recognition to maintain a given steady-state recognition $R$ (Appendix~\ref{app_C}). The effective number of adaptive paths is determined solely by the genetic architecture of the recognition trait; the actual speed of adaptation, or fitness flux, also depends on the global parameters $\mu$, $N$, $\ep_0$, and on the fitness amplitude $f_0$~\cite{mustonen2010fitness}. In the minimal model, the ML number of adaptive paths is related in a simple way to coding density and code length. When code changes are under strong selection, we find  
\EQ
A^* = \frac{\gamma_0}{1 - \gamma_0} \,  (1 - \gamma^*) \, \ell^*
\label{A}
\EE 
(Appendix~\ref{app_C}). In the strong-driving regime, $A^*$ is a strongly increasing function of the driving rate (Fig.~\ref{fig:fig4}F). An increased value of $A$ implies a higher supply of beneficial mutations and faster response to moving targets. 
We have shown that by increasing molecular complexity, adaptive ratchets actually evolve increased values of $A$ depending on the driving rate: the recognition system adapts its own adaptability to recognize moving targets.

\subsection*{Adaptive tinkering and functional alterations} 
Around the strong-driving ML point $\ell^*$, the ratchet dynamics of the code length becomes diffusive ($|v_e | \ll v_{\rm tot}$), as in near-neutral evolution, but the diffusion rate is 
adaptively enhanced ($v_{\rm tot} \gg 2 \nu$, Fig.~\ref{fig:fig4}F). These characteristics define a ratchet regime of adaptive tinkering at large code length. The range of this regime is given by the width of the potential peak $\Psi(\ell)$, which is defined as the r.m.s. variation $\Delta \ell$ around $\ell^*$ and can be computed from the curvature of the evolutionary potential. We find a broad regime of adaptive tinkering, $\Delta \ell \sim \ell^*$ with a proportionality factor of order 1 (Fig.~\ref{fig:fig5}D, Appendix~\ref{app_C}). Rapid diffusion and broad length distributions are specific effects of adaptive ratchets. For evolution in a single-valued fitness landscape $F (\ell) = F_r (\ell) - c_0 \ell$, the diffusion rate at the peak is bounded by the neutral rate, $v^*_{\rm tot}  \leq 2 \nu$. Moreover, if $F_r$ is a power of $\ell$, length fluctuations around the peak scale as $\Delta \ell \sim (\ell^*)^{1/2}$ (Appendix~\ref{app_C}). 

Together, the turnover of recognition code includes point mutations at rates $u_+^* \sim \kappa \mu$ and length  changes at rates $v^*_{\rm tot}  \sim \kappa \nu$, both of which are tuned by the seascape of target changes. Thus, by adaptive tinkering, recognition sites rapidly scan sequence space while their function is maintained. This enhances the probability of evolving functional refinements (for example, more specific target binding by epistatic interactions between sequence units) and of adopting new functions (examples are overlapping binding sites in bacteria~\cite{kovacikova2001overlapping,haycocks2016unusually}). Evolutionary paths of recognition sites can also end with loss of function. (Appendix~C, Fig.~C2). Short functional sites can be lost because they cannot keep up with a moving target of high $\kappa$; for longer sites, entropic forces and the cost of complexity can conspire to decrease $\Delta G$ and $\ell$. We conclude that adaptive ratchets evolve a transient gain of molecular complexity, which can be locked in by additional functions or is exposed to pruning.

\begin{figure*}[t!]
\centering
\includegraphics[width=0.85\textwidth]{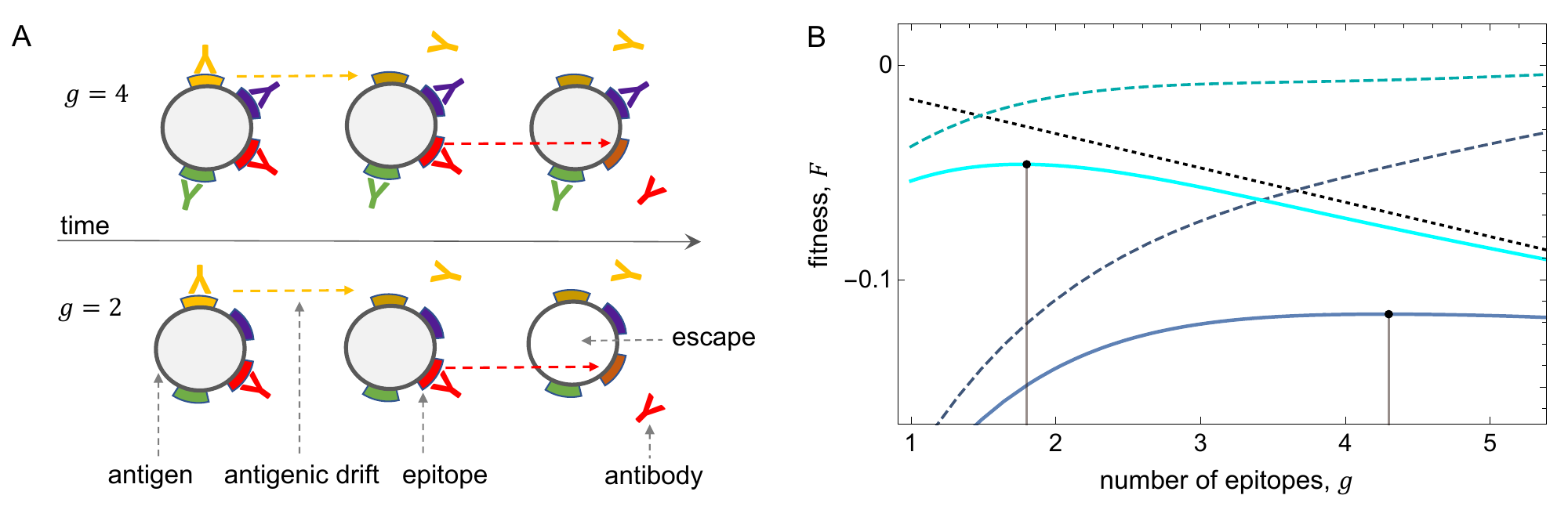}
\caption{{\bf Complexity-dependent recognition and fitness of the adaptive immune system}. 
(A) Immune recognition and escape evolution (schematic). Antibodies neutralize pathogens by binding to specific target sites (epitopes). The number of initially bound epitopes, $g$, measures the complexity of the immune response. Antigenic evolution of the pathogen (dashed arrows) changes the target sequences and reduces recognition of each epitope, eventually leading to escape from immune recognition. These dynamics is shown for $g = 2$ and $g = 4$. 
(B) Complexity-dependent fitness landscape, $F (g)$ (solid), recognition component, $F_r (g)$ (dashed), and complexity cost, $F_c (g)$ (dotted black). ML values $g^*$ are marked by black dots; $F_r (g)$ is plotted relative to the maximum recognition fitness and in units of the infection cost parameter $f_0$. 
This landscape is shown for slow- and fast-evolving pathogens with effective antigenic mutation rates $\kappa = 2$ (blue) and $\kappa = 0.08$ (cyan). To be compared with Fig.~5AC. Evolutionary parameters: $\ep_0 = 2.5$, $ \Delta G_i=\Delta G_{50} + 2\ep_0 k_BT$,  $c_0 / f_0 = 0.016$. 
  }
  \label{fig:fig6}
\end{figure*}

\section{Complexity of immune recognition}
The following example shows how selection on complexity can act in the adaptive immune system, a rapidly evolving recognition system of high global complexity~\cite{altan2020quantitative}. In the presence of an antigen, immune B cells produce neutralizing antibodies that bind to specific target sites on its surface (called antigenic epitope sites)~\cite{lam2019multifaceted}. In a primary infection, a part of the responding B cells is stored as immune memory to protect against future infections by the same pathogen (another part evolves high affinity to their cognate epitope by affinity maturation, a rapid evolutionary process under selection for recognition~\cite{eisen1964affinitymaturation, victora2012germinalcenters,viant2020memory,glaros2021earlymemory}). Fast-evolving pathogens, however, are moving targets: they change epitope sequences by accumulation of mutations, which can lead to eventual escape from immune recognition and protection. This process, called antigenic drift, is frequently observed in RNA viruses, including human influenza, norovirus, and SARS-CoV-2~\cite{earn2002ecology, smith2004mapping, white2014evolution, meijers2023populationimmunity}. Here we develop a minimal biophysical model for the immune recognition of an evolving antigen and its consequences for host fitness. 

A natural measure for the molecular complexity of immune responses to a given antigen is the effective number of epitopes targeted by neutralizing antibodies. For an antigen with a total of $g$ epitopes, we record the epitope-specific antiserum potencies, $\Z = (Z_1, \dots, Z_g)$, which measure the relative probabilities that each epitope is bound by antibodies~\cite{morantovar2024primaryresponse}. We then define the immune recognition complexity $g_r$ in terms of a Shannon entropy, $\log g_r \equiv \langle S(\z) \rangle$, where $\z = \Z/ \sum_{\alpha=1}^g Z_\alpha$ is the potency distribution of a given host and angular brackets denote averages over the host population. Immune recognition complexity is closely related to the established notion of immunodominance~\cite{abbott2020immunodominance, ertl2003immunodominant} -- low complexity implies high immunodominance and vice versa -- and is known to vary substantially between viral pathogens and between human hosts~\cite{childs2015trade, lee2019mapping, munoz2021serotypic}. 

The complexity of immune recognition is a co-evolutionary trait: it depends on pathogen factors, such as the number of accessible epitopes, and 
on global parameters of the host immune repertoire, including the number of B cell lineages and the length of the antibody-epitope binding motif~\cite{morantovar2024primaryresponse}; more details are given in Appendix~\ref{app_D}. Therefore, hosts can change the effective number of targeted epitopes by long-term evolution of their immune systems. For example, humans and mice differ substantially in the number of B cell lineages and consistently in the immunodominance of B cell immune responses~\cite{glanville2009diversity, elhanati2015diversity}. 

\subsection*{Model for recognition function} 
Infections by an antigen with $g$ accessible epitopes induce 
the production of an antiserum with epitope-specific potencies $\Z$. We define the effective target affinities $\log \Z \equiv (\Delta G_1, \dots, \Delta G_g) / (k_BT)$ as the biophysical recognition phenotype for the immune memory response of this system. These affinities govern the recognition of each epitope, $R_\alpha = H (\Delta G_\alpha - \Delta G_{50})$, where $\Delta G_{50}$ denotes a threshold antibody level required for clinical protection against re-infection~\cite{hobson1972titersprotection, coudeville2010titersprotection}. The form of the single-epitope recognition function is analogous to Eq.~(\ref{R_min}) in our model for gene regulation. The full protection function takes the form $R = 1 - (1 - R_1) \times \dots \times (1 - R_g)$, assuming an exposed individual is protected as long as the antigen is recognized at any of its epitopes. 

This recognition function is well supported by empirical findings. First, clinical studies show that human immune protection correlates well with overall antibody titers in serum, independently of the epitope distribution of the response~\cite{hobson1972titersprotection, coudeville2010titersprotection}. Second, functional binding of a single epitope (monoclonal response) is sufficient to confer full protection to the host~\cite{lee2019mapping}; consistently, a single epitope mutation can cause loss of protection~\cite{koel2013singlemutation}. 

\subsection*{Complexity-dependence of host fitness} 
Next, we describe how the immune recognition in a secondary infection changes by antigenic drift of the pathogen. A minimal molecular model of this process is sketched in Fig.~\ref{fig:fig6}. An infection generates immune memory with $g$ equal high-affinity components $\Delta G_\alpha = \Delta G_0 > G_{50}$ ($\alpha = 1, \dots, g$) against the infecting strain, resulting in a recognition complexity $g_r = g$. Subsequent antigenic drift occurs by epitope mutations with effect $\Delta \Delta G_\alpha / (k_B T) = \ep_0$ and point substitution rate $\rho$, generating stochastically reduced affinities $\Delta G_\alpha$ against evolved strains. Thus, antigen evolution degrades the expected molecular recognition $R (t)$ at a time $t$ after the previous infection, leading eventually to loss of recognition and re-infection. We evaluate the stationary-state population average of recognition, $\langle R \rangle$, describing a population of adult individuals subject to multiple re-infections by evolving pathogens. 

The joint dynamics of pathogen evolution and immune adaptation generates fitness effects for the host system. We study a minimal fitness landscape $F(g)$ describing the dependence of host fitness on the molecular complexity of immune recognition. A recognition-dependent term $F_r (g) = f_0 \langle R \rangle$ is generated by the fitness cost of infections. A fitness cost of complexity, $F_c (g)$, can arise from multiple factors, including the size of the naive immune repertoire: larger recognition complexity $g$ requires a larger number of naive B cell lineages~\cite{morantovar2024primaryresponse}, generating a larger burden of repertoire generation, maintenance, and off-target recognition (Appendix~\ref{app_D}). For simplicity, we summarize these costs into a linear fitness term, $F_c (g) = - c_0 g$. The resulting minimal model for host fitness takes the form $F (g) = F_r (g) + F_c (g)$, which is analogous to Eq.~[\ref{F}] in our evolutionary model for gene regulation. 



Fig.~\ref{fig:fig6}B shows the population average of the steady-state fitness as a function of the number of targeted epitopes, $g$, for different values of the effective speed of antigenic evolution, $\kappa = \rho \tau$. We find a link between immune recognition complexity and target mutation rate remarkably similar to Fig.~\ref{fig:fig5}AC: multi-epitope immune responses play out a fitness benefit $F_r (g)$ specifically against fast-evolving antigens ($\kappa \gtrsim 1$). Again, the recognition fitness component shows a diminishing-return nonlinearity, leading to a $\kappa$-dependent maximum of the total fitness $F(g)$. The underlying dynamical mechanism is similar as well: immune responses targeting complex epitope states adapt more efficiently to moving antigenic targets. A related effect has recently been discussed in ref.~\cite{sachdeva2020tuning}: antigenic drift can favor generalist over specialist immune responses against multiple antigens. For even faster target evolution ($\kappa \gg 1$), defense by immune memory becomes inefficient, in accordance with the results of ref.~\cite{Mayer2019}. 

In summary, we have shown that fast antigenic drift can induce selection for complex adaptive immune responses, in line with our general picture  of complexity in dynamical recognition systems. This finding 
is consistent with observations: several fast-evolving RNA viruses, including influenza~\cite{wilson1990structural, klingen2018silico}, norovirus~\cite{white2014evolution}, and Sars-Cov-2~\cite{barnes2020sars, greaney2021mapping, harvey2021sars}, are subject to multi-epitope responses, although successful targeting of a single epitope is sufficient for neutralization~\cite{lee2019mapping}. 
However, the minimal model is insufficient to describe the full dynamics of immune recognition complexity. These processes depend on the long-term co-evolution of host immune systems and multiple viral pathogens \cite{chardes2022coevolution}, which is beyond the scope of this paper. Other features of immune recognition not contained in the minimal model include a feedback of recognition onto the speed of target evolution. If epitope mutations come with a collateral cost to the pathogen (for example, through their effect on protein stability), a complex immune response can slow down or halt antigenic drift; this effect has recently been reported for measles~\cite{munoz2021serotypic}.

\section*{Discussion} 

Organisms live in dynamic environments. Changing external signals continuously degrade the fidelity of an organism's recognition units and generate selective pressure for change. Here we argue that stochastic, adaptive evolution of molecular interactions drives the complexity of the underlying sequence codes. By analytical computation and {\em in silico} evolution experiments, we show that recognition sites tuned to a dynamic molecular target evolve a larger code length than the minimum length, or algorithmic complexity, required for function (Fig.~3). The increase of coding length is coupled to a decrease of coding density; that is, individual letters carry a reduced information about the target. Importantly, these shifts also increase the number of mutational paths available for adaptation of recognition and the overall turnover rate of the recognition site sequence. Complex sites evolve in a specific regime of adaptive tinkering, which facilitates the maintenance of recognition interactions in dynamic environments. Moreover, the rapid parsing of fuzzy recognition site sequences increases the propensity for functional refinement or alteration, for example by tuning epistatic interactions within the recognition site or by recruiting additional targets for cooperative binding. We can summarize these dynamics as an evolutionary feedback loop: moving targets, by increasing the molecular complexity of their cognate recognition machinery, improve their own recognition.

Our analysis suggests specific ratchet modes of evolution by which recognition sites gain molecular complexity (Fig.~5). All of these evolutionary ratchets build on a separation of mutational time scales between functional changes at fixed complexity and complexity changes. This is a natural assumption in general, because alterations in the architecture of recognition interfaces are subject to stronger conformational and selective constraints than  point mutations that tune binding affinities. We have shown that the separation of time scales generates a fundamental asymmetry: code length extensions take place into random sequence, compression steps cut into functional sequence, causing asymmetric selection for reverse evolutionary steps. Therefore, even without any explicit fitness benefit, complexity can evolve as a collateral of selection for function. 

We find two modes of complexity evolution, which are distinguished by the selective forces governing the ratchet. A classical ratchet operates by neutral steps of complexity increase and selective constraint against reverse steps. A neutral path to higher complexity can be realized under two conditions: the cost of complexity is small and the primary function is well-adapted. In this mode, negative selection against loss of complexity builds up as a collateral of stabilizing selection on the primary function in the complex state (Fig.~5B). For example, in Lynch and Conery's model of macro-evolution, the genomes of higher organisms grow neutrally because reduced effective population sizes reduce the cost of complexity; later, functions can latch on and eventually be conserved by selective constraint~\cite{lynch2003origins}. Similarly, in entrenchment models of protein evolution, complex multi-domain states establish neutrally if the cost of formation is small. Such complexes can be maintained by an accumulation of mutations that are conditionally neutral but deleterious in simpler states~\cite{hochberg2020hydrophobic}. 

In contrast, adaptive ratchets operate in a time-dependent environment that generates a fitness seascape: external changes drive the system away from its fitness peak and are balanced by continuous adaptive evolution. Adaptive ratchets generate positive selection on complexity changes coupled to functional adaptation; i.e, as a collateral to positive selection on the primary function (Fig.~5D). Importantly, this mode of evolution can override a substantial cost of complexity. The rate of adaptive ratchets is set by external driving forces, here recognition target changes, not by genetic drift. Hence, unlike classical arguments of complexity evolution, adaptive ratchets do not describe the refinement of functions in a static environment but the maintenance of functions in the face of a fitness seascape.

How universal are evolutionary ratchets? We can expect this mechanism to operate in a broad range of molecular networks with multiple parallel sub-units and a reservoir of potential units. Examples include recognition sites with multiple sequence matches, repertoires of immune response with multiple antibody lineages, and on larger scales, regulatory or metabolic networks with multiple genes contributing to the same function. Such networks allow ratchet modes of evolution to kick in: by preferential attachment of fit and deletion of unfit sub-units, changes in network architecture can couple to selection for network function. Specifically, the adaptive ratchet model then predicts that external driving -- here by moving recognition targets, evolving antigens, or time-dependent metabolic substrates -- induces the evolution of complex networks with redundant sub-units. 

New high-throughput methods are currently used to map sequence and phenotypes of molecular interactions at an unprecedented depth. Prominent examples include transcription factor  interactions in microbes~\cite{kinney2010using,barnes2019mapping,ireland2020deciphering} and receptor-antigen interactions in human immune repertoires~\cite{doud2017complete,marcou2018high,minervina2020primary}. This opens the perspective of new experimental screens to map correlations between external driving, network architecture, and molecular complexity. Experiments with exposure to moving targets can determine in which mode -- constraint or adaptation -- ratchet evolution is determining the complexity of these systems. Such experiments can also become a powerful tool for synthetic evolution, to elicit molecular complexity of recognition and its potential for evolutionary innovation. 

\section*{Acknowledgments} We thank Rob Phillips for discussions. This work was supported by Deutsche Forschungsgemeinschaft grant CRC 1310 ``Predictability in Evolution''. 

\subsection*{Author Contributions} All authors performed model development, analytical calculations, and numerical simulations. TR focused on the general model, RMT on the application to immune systems. All authors contributed to writing the paper.

\clearpage

\appendix
\counterwithin{figure}{section}

\section{Evolutionary model}
\label{app_A}

\subsection{Recognition trait and function}
In a biophysical model of recognition, a quantitative trait characterising a recognition site is its binding affinity to the recognition target, $\Delta G$. Here we measure this quantity with the zero point given by its expectation value in random sequence; the results do not depend on this choice. The recognition function, defined as the equilibrium occupancy of a binding site with affinity $\Delta G$ at physiological temperature $T$ is then given by 
\EQ
R = \frac{1}{1 + \exp[(\Delta G_{50} - \Delta G (\gamma, \ell)) / k_B T]}, 
\label{Hill} 
\EE
where $\Delta G_{50}$ denotes the functionality threshold (half-binding point). In our minimal model, the reduced binding affinity takes the simple form 
\EQ
\frac{\Delta G (\gamma, \ell)}{k_B T} = (\gamma - \gamma_0) \ell \ep_0, 
\label{DeltaG_min} 
\EE
which depends on the site length $\ell$, the fraction of target-matching sequence units in the site, $\gamma = k/\ell$, the average fraction of matches in random sequence, $\gamma_0$, and the reduced affinity gain per match, $\Delta \Delta G / k_B T = \ep_0$. 

For a given background sequence $\a = (a_1, \dots, a_L)$, the functionality threshold is given by the condition that the Boltzmann factor of the functional site equals the sum of Boltzmann factors for spurious binding at all background sequence positions, 
\EQ
N \exp \left [ \frac{\Delta G_{\rm 50} (\a)}{k_B T} \right ]   = Z_b (\a) 
\label{Zb} 
\EE
with 
\EQ
Z_b (\a) = \sum_{r = 1}^{L - \ell +1} \exp \left [ \frac{\Delta G (a_r, \dots, a_{r+\ell -1}) - \Delta G_0}{k_B T} \right ] .  
\EE
Here $N$ is the expected number of TF molecules diffusing along the DNA backbone, and $\Delta G_0$ is the affinity of the diffusive molecular state. A typical transcription factor in prokaryotes has a 1D diffusion density $\rho = N/ L \ll 1$, which is a finite fraction of the total TF density in the cell \cite{mirny2009protein}.

\begin{figure}[h!]
  \begin{center}
  \includegraphics[width=\linewidth]{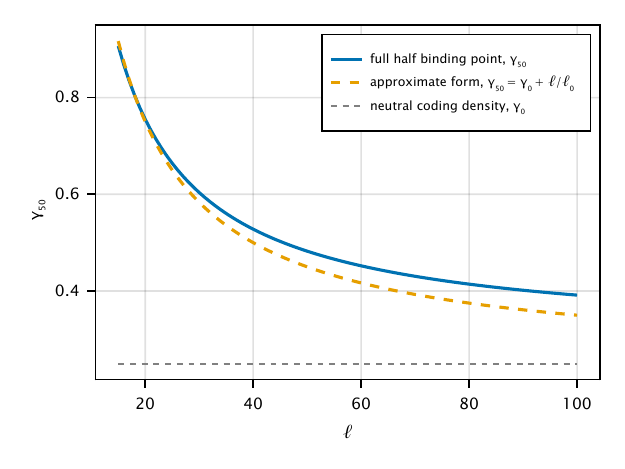}
  \end{center}
  
  \caption{
  {\small {\bf Length dependence of the recognition threshold.} 
The half-binding coding density $\gamma_{50}$ of the minimal recognition model is shown in the full form (\ref{gamma50}) (blue) and in the approximation used for analytical estimates, $\gamma_{50} = \gamma_0 + \ell_0/\ell$ (orange). The neutral coding density $\gamma_0$ (gray) is shown for comparison. Parameters as in Fig.~2. 
}
 }
  \label{fig:half_binding}
  \end{figure}

The expectation value of the functionality threshold in a random ensemble of background sequences is given by the quenched average 
\EQ
\frac{\Delta G_{\rm 50}}{k_B T} = - \log N + \sum_\a P_0 (\a) \log Z_b (\a), 
\label{DeltaG_q}
\EE
where $P_0 (\a)$ denotes the probability of genotype $\a$ in the background ensemble. Following ref.~\cite{gerland2002physical}, we evaluate Eq.~(\ref{DeltaG_q}) in the so-called annealed approximation, averaging over $Z_b$ instead of $\log Z_b$. In this approximation, the average factorizes into the contributions of individual sequence positions, 
\EQA
\frac{\Delta G_{\rm 50}}{k_B T} & = & - \log N + \log \sum_\a P_0 (\a) Z_b (\a)
\nonumber \\
& = & - \log N + \log \left [ L \prod_{i=1}^\ell \sum_a p_0 (a) \, {\rm e}^{\ep_i (a)} \right ] , 
\EEA
where $p_{0} (a)$ denotes the probability of nucleotide $a$ and $\ep_i (a)$ its affinity contribution at position $i$ of the site. In the minimal model with our convention on the zero point of affinity, matches occur with probability $p_{0,+} = \gamma_0$ and have $\ep_+ = (1 - \gamma_0) \ep_0$, mismatches  occur with probability $p_{0,-} = 1 - \gamma_0$ and have $\ep_- = - \gamma_0 \ep_0$. We obtain 
\EQ
\frac{\Delta G_{\rm 50}}{k_B T} =-\log \rho  + \ell \log \left[ \gamma_0 {\rm e}^{(1 - \gamma_0) \, \ep_0} +  
(1 - \gamma_0) \, {\rm e}^{- \gamma_0 \ep_0} \right].
\EE
The first term depends on the density of TF molecules but is independent of $\ell$; the second term accounts for the sequence heterogeneity of the random ensemble. As dicussed in the main text, TF binding sites have a minimum functionality threshold $\Delta G_{50}^{\min}$, corresponding to a minimum number $\ell_0 = \Delta G_{50}^{\min} / (\ep_0 k_B T)$ of excess matches over random sequence and a minimum length $\ell_{\min} = \Delta G_{50}^{\min} / (\ep_0 (1 - \gamma_0) \, k_B T)$. Thus, we we can rewrite the functionality threshold in the form 
\EQ
\frac{\Delta G_{\rm 50}}{k_B T} = \epsilon_0 \ell_0  + (\ell - \ell_{\min}) \frac{\Delta \Delta g_{50}}{k_B T} 
\label{DeltaG50} 
\EE
with a threshold shift per unit length,
\EQ
 \frac{\Delta \Delta g_{50}}{k_B T} = 
  \log \left[ \gamma_0 {\rm e}^{(1 - \gamma_0) \, \ep_0} +  
(1 - \gamma_0) \, {\rm e}^{- \gamma_0 \ep_0} \right] . 
\EE
By equation~(\ref{DeltaG_min}), the half-binding point corresponds to a code density
 \EQA
\gamma_{\rm 50} & = & \gamma_0 + \frac{1}{\ep_0 \ell} \frac{\Delta G_{\rm 50}}{k_B T}
\nonumber \\
& = &  \gamma_0 + \frac{\ell_0}{\ell} + \frac{(\ell - \ell_{\min})}{\epsilon_0 \ell} \frac{\Delta \Delta g_{50}}{k_B T}. 
\label{gamma50} 
\EEA
In our numerical analysis reported in Figs.~3-5 and 7-11, we use a recognition function (\ref{Hill}) with the full, length-dependent form (\ref{DeltaG50}) -- (\ref{gamma50}) of the recognition threshold. The length-dependent term remains subleading over the range of relevant coding lengths, as shown in Fig.~\ref{fig:half_binding}. 
Therefore, we use the length-independent form $ \Delta G_{50} / (k_B T) \approx \epsilon_0 \ell_0$, corresponding to 
$\gamma_{50} \approx \gamma_0 + \ell_0 /\ell$ in the analytical approximations, as in equation~(\ref{expapprox}).

\subsection{Sequence evolution rates} 
In the minimal model, the sequence evolution of recognition involves the three classes of changes shown in Fig.~1. (i) Point mutations of the recognition sequence at constant code length occur at a rate $\mu$ per unit of sequence (Fig.~1A). These changes have point mutation rates 
\EQA
\mu_+ (\gamma) & = & \mu (1 - \gamma) \frac{\gamma_0}{1 - \gamma_0}
\nonumber \\
\mu_- (\gamma) & = & \mu \gamma,  
\label{mupm}
\EEA
where the mutation rates are weighted such they only reflect mutations that lead to changes in coding density. These mutations have selection coefficients
\EQA
s_+ (\gamma, \ell) & = & s_\gamma (\gamma + \mbox{$\frac{1}{2\ell}$}, \ell)
\nonumber \\
s_- (\gamma, \ell) & = & - s_\gamma (\gamma - \mbox{$\frac{1}{2\ell}$}, \ell).
\label{spm}
\EEA
depending on the direction of selection (beneficial/deleterious, indicated by the index $\pm$). Here  
\EQ
s_\gamma (\gamma, \ell) = F (\gamma + \mbox{$\frac{1}{2\ell}$}, \ell) - F(\gamma - \mbox{$\frac{1}{2\ell}$}, \ell) 
\label{s_gamma}
\EE
is the fitness increment in the landscape (1) in midpoint discretisation. Eqs.~(\ref{mupm}) and~(\ref{spm}) determine the substitution rates 
\EQA
u_+ (\gamma, \ell) & = & \mu_+ (\gamma) \, g(s_+ (\gamma,  \ell))
\nonumber \\
u_- (\gamma, \ell) & = & \mu_- (\gamma) \, g(s_- (\gamma,  \ell))
\label{upm}
\EEA
in the low-mutation regime ($\mu N \ell \lesssim 1$). Here $g(s) = N \pi (s)$ is Haldane's factor in a population of effective size $N$ \cite{kimura1962probability}, 
\EQ
g (s) = \frac{2Ns}{1 - \exp(-2N s)}, 
\label{haldane} 
\EE
which is proportional to the fixation probability of an individual mutation, $\pi (s)$. With an appropriately changed effective population size, these substitution rates apply even if the global sequence evolution takes place under clonal interference \cite{good2014genetic,held2019survival}. 
(ii)~Recognition site extensions and compressions add or remove one unit of sequence (Fig.~2BC) and occur at a rate $\nu \ll \mu$. These changes have mutation rates 
\EQA
\nu_{++} (\gamma) & = & \nu \gamma_0
\nonumber  \\
\nu_{+-} (\gamma) & = & \nu (1 - \gamma_0)  
\nonumber  \\
\nu_{-+} (\gamma) & = & \nu (1 - \gamma)  
\nonumber  \\
\nu_{--} (\gamma) & = & \nu \gamma  
\label{nupmpm}
\EEA
and selection coefficients
\EQA
s_{++} (\gamma, \ell) & = & F \big (\mbox{$\frac{\gamma + 1/\ell}{1 + 1/\ell}$} , \ell +1 \big ) - F(\gamma, \ell) 
\nonumber \\
s_{+-} (\gamma, \ell) & = & F \big (\mbox{$\frac{\gamma }{1 + 1/\ell}$} , \ell +1 \big ) - F(\gamma, \ell)
\nonumber \\
s_{-+}(\gamma, \ell)  & = & F \big (\mbox{$\frac{\gamma }{1 - 1/\ell}$} , \ell -1 \big ) - F(\gamma, \ell)  
\nonumber \\
s_{--} (\gamma, \ell) & = & F \big (\mbox{$\frac{\gamma - 1/\ell}{1 + 1/\ell}$} , \ell -1 \big ) - F(\gamma, \ell).
\label{spmpm}
\EEA
For the subsequent analytical treatment, we use the exponential approximation of the target binding probability (\ref{Hill}), (\ref{DeltaG_min}) in the regime of functional sites,
\EQA
R(\gamma,\ell) & \simeq & 1 - \exp \left [\frac{\Delta G_{50}}{k_BT} - (\gamma - \gamma_0) \ell \ep_0 \right ] 
\nonumber \\
& \approx & 1 - \exp\left[\epsilon_0\ell_0 - \epsilon_0\ell(\gamma - \gamma_0) \right ]. 
\label{expapprox}
\\ \nonumber 
\EEA
In the last expression, we also use the length-independent form $\Delta G_{50}/(k_BT) \approx \ep_0 \ell_0$, which provides a good approximation to the full form (\ref{DeltaG50}) throughout the range of site lengths relevant for this paper (Fig.~3, Fig.~\ref{fig:half_binding}). 

We can then relate the selection coefficients (\ref{spmpm}) to the selection coefficient for point mutations of the recognition sequence in Eq.~(\ref{s_gamma}),
\EQA
s_{++}(\gamma,\ell) & = &c_{++} s_\gamma(\gamma,\ell)  - c_0
\nonumber\\
s_{+-}(\gamma,\ell) & = &c_{+-} s_\gamma(\gamma,\ell) - c_0
\nonumber\\
s_{-+}(\gamma,\ell) & = &  c_{-+} s_\gamma(\gamma,\ell) + c_0
\nonumber\\
s_{--}(\gamma,\ell) & = & c_{--}  s_\gamma(\gamma,\ell) + c_0, 
\EEA
where $c_0$ denotes the complexity cost parameter in the fitness landscape Eq.~(1) and 
\EQA
c_{++} & = & \frac{1-\exp\left[-\epsilon_0(1-\gamma_0)\right]}{2\sinh(\epsilon_0/2)},
\nonumber\\
c_{+-} & = & \frac{1-\exp\left[\epsilon_0\gamma_0\right]}{2\sinh(\epsilon_0/2)},
\nonumber\\
c_{-+} & = & \frac{1-\exp\left[-\epsilon_0\gamma_0\right]}{2\sinh(\epsilon_0/2)},
\nonumber\\
c_{--} & = & \frac{1-\exp\left[\epsilon_0(1-\gamma_0)\right]}{2\sinh(\epsilon_0/2)}. 
\EEA
Here we compute one selection coefficient explicitly, 
\begin{widetext}
\EQA
s_{++} (\gamma, \ell) & = & F \big (\mbox{$\frac{\gamma + 1/\ell}{1 + 1/\ell}$} , \ell +1 \big ) - F(\gamma, \ell)
\nonumber \\[0.8em]
&=& f_0\left(\exp\left[ \epsilon_0\ell_0  - \ell(\gamma - \gamma_0)\right] - \exp\left[ \epsilon_0\ell_0 - (\ell+1)\left(\frac{\gamma+1/l}{1+1/l} - \gamma_0\right)\right]\right) - c_0
\nonumber \\[0.8em]
&=& f_0\left(\exp\left[ \epsilon_0\ell_0  - \epsilon_0\ell(\gamma - \gamma_0)\right] - \exp\left[ \epsilon_0\ell_0 - \epsilon_0\ell\left(\gamma - \gamma_0\right) -\epsilon_0 + \epsilon_0\gamma_0\right]\right) - c_0
\nonumber \\[0.8em]
&=&  f_0\exp\left[ \epsilon_0\ell_0  - \epsilon_0\ell(\gamma - \gamma_0)\right]\left(1 - \exp\left[ -\epsilon_0(1-\gamma_0) \right]\right) - c_0
\nonumber \\[0.8em]
&=& f_0 (\exp[\epsilon/2] - \exp[-\epsilon/2] ) \exp\left[ \epsilon_0\ell_0  - \epsilon_0\ell(\gamma - \gamma_0)\right] c_{++} - c_0
\nonumber \\[0.8em]
&=& f_0\left(\exp\left[ \epsilon_0 \ell_0  - \epsilon_0 \ell(\gamma - 1/2l - \gamma_0)\right] - \exp\left[ \epsilon_0 \ell_0  - \epsilon_0 \ell(\gamma + 1/2l - \gamma_0)\right]\right) \, c_{++} - c_0
\nonumber \\[0.8em]
&=& c_{++} s_\gamma(\gamma,\ell) - c_0; 
\EEA
\end{widetext}
the other coefficients are obtained in the same way. These selection coefficients depend on the direction of change (extension/compression, first index $\pm$) and the direction of selection (beneficial/deleterious, second index $\pm$).  The approximate expressions in terms of the fitness increment $s_\gamma$ are used for the analytical closure discussed below. The exponential prefactors, which are of order 1, reflect the shift in the half-binding point $\gamma_{50}$ induced by a length change and the switch to the midpoint discretization in Eq.~(\ref{s_gamma}). 
Eqs.~(\ref{nupmpm}) and~(\ref{spmpm}) determine the total substitution rates of extensions and compressions, 
\EQA
v_{+} (\gamma, \ell) & = &  \nu_{++} (\gamma) \, g(s_{++} (\gamma, \ell)) + \nu_{+-} (\gamma)  \, g(s_{+-} (\gamma, \ell))
\nonumber \\
v_{-} (\gamma, \ell) & = & \nu_{-+} (\gamma)  \, g(s_{-+} (\gamma, \ell)) + \nu_{--} (\gamma)  \, g(s_{--} (\gamma, \ell)), 
\nonumber \\
\label{vpm} 
\EEA
(iii)~Recognition target changes occur at a rate $\rho = \kappa \mu$ per unit of sequence, independently of the recognition function (Fig.~1D). Hence, beneficial and deleterious target changes have rates
\EQA
\rho_+ (\gamma) & = & \kappa \mu (1 - \gamma) \frac{\gamma_0}{1 - \gamma_0}
\nonumber \\
\rho_- (\gamma) & = & \kappa \mu \gamma,  
\label{rhopm}
\EEA
which have the same combinatorial factors as the point mutations (\ref{mupm}) and are independent of the fitness landscape $F(\gamma, \ell)$. 

Individual evolutionary trajectories $(k, \ell)(t)$ generated by the substitution dynamics with rates $u_\pm (\gamma, \ell)$, $v_\pm (\gamma, \ell)$, and $\rho_\pm (\gamma)$ are shown in Fig.~2BD and in Appendix C, Fig.~\ref{fig:loss_traj}.\\

\section{Stationary-state analysis} 
\label{app_B}

\subsection{Evolutionary potentials and stationary distribution} 

Given the separation of mutational time scales for coding density and code length ($\nu \ll \mu$), the stationary distribution factorizes, 
\EQ
Q(\gamma, \ell) = Q(\gamma | \ell) \, Q(\ell). 
\label{Q}
\EE
The conditional distribution of the coding density takes the form 
\EQ
Q(\gamma | \ell) = \exp[\Psi(\gamma | \ell)]
\EE
with a potential function given by detailed balance, 
\EQA
\Psi (\gamma | \ell) & = & \sum_{k'=0}^{\gamma \ell} \log \frac{u_+ (k'/\ell, \ell) + \rho_+ (k'/\ell, \ell)}{u_- ((k'+1)/\ell, \ell) + \rho_- ((k'+1)/\ell, \ell)}  
\nonumber \\
& & - \log Z_\gamma (\ell).
\label{Psi_gamma}
\EEA
Here $Z_\gamma (\ell)$ is a normalization factor and the point substitution rates $u_\pm, \rho_\pm$ are given by Eqs.~(\ref{upm}) and~(\ref{rhopm}). This distribution defines the conditional average of an observable $x(\gamma, \ell)$, 
\EQ
\bar x (\ell) \equiv \sum_\gamma x (\gamma, \ell) \, Q(\gamma | \ell). 
\label{Xbar} 
\EE 
Because $Q(\gamma | \ell)$ is approximately Gaussian (Fig.~B1), such averages can be accurately evaluated by steepest descent, $\bar x (\ell) \approx x (\gamma^*(\ell), \ell)$, where $\gamma^*(\ell) = \arg \max_\gamma Q(\gamma | \ell)$ denotes the conditional maximum likelihood (ML) coding density. In particular, this procedure yields an effective fitness landscape for code length, 
\EQ
\bar F (\ell) = \bar F_r (\ell) + F_c (\ell) = f_0 - \frac{1}{\ep_0} s_\gamma^* (\ell) - c_0 \ell,
\label{Fbar}
\EE
where $s_\gamma^* (\ell) \equiv s(\gamma^*(\ell), \ell)$ is the selection coefficient at the ML coding density. Here we use an exponential approximation of the Hill recognition function $R(\gamma, \ell)$ in the regime of functional sites, which implies $s_\gamma = \ep_0 \partial F_r / \partial (\Delta G / k_B T) = \ep_0 (f_0 - F_r)$. 
The marginal distribution of code length, 
\EQ
Q(\ell) = \exp[\Psi(\ell)], 
\EE
is again determined by detailed balance, 
\EQ
\Psi (\ell) = \sum_{\ell'=\ell_\mathrm{min}+1}^{\ell} \log \frac{\bar v_+ (\ell' - 1)}{\bar v_- (\ell')} - \log Z_\ell, 
\label{Psi_ell}
\EE
where $Z_\ell$ is a normalization factor, $\ell_\mathrm{min}$ is the minimal length of functional sites and $\bar v_\pm$ are the average extension and deletion rates given by Eqs.~(\ref{vpm}) and~(\ref{Xbar}). From the full distribution $Q(\gamma, \ell)$, we define the corresponding distribution for functional sites, 
\EQ
Q_f (\gamma, \ell) =  \frac{1}{Z_f} \, R(\gamma, \ell) \, \exp[\Psi (\gamma, \ell)] . 
\label{Q_f} 
\EE
This distribution weighs sequence states by the recognition (target binding) probability, $R(\gamma, \ell)$. Examples of the distribution of functional sites $Q_f (\gamma, \ell)$ are plotted in Fig.~2AC; the evolutionary potential $\Psi (\ell)$ is shown in Fig.~4BD. \\

\subsection{Weak-driving regime} 
For $\kappa \lesssim 1$, the evolutionary potential for coding density takes the form 
\EQ
\Psi (\gamma | \ell) \simeq S (\gamma, \ell) + \frac{2 N}{(1 + \kappa)} \, F_r (\gamma, \ell) - \log Z_\gamma (\ell), 
\label{Psi_gamma_QG}
\EE
where 
\EQA
S(\gamma, \ell) & = & \log \left [ \binom{\ell}{\gamma \ell} \gamma_0^{\gamma \ell} (1 - \gamma_0)^{(1 - \gamma)\ell} \right ] 
\nonumber \\
& = & {\rm const.} - \frac{\ell \, (\gamma - \gamma_0)^2}{2 \gamma_0 (1 - \gamma_0)} + O((\gamma - \gamma_0)^3)
\nonumber \\
& & 
\label{S} 
\EEA
is the entropy, defined as the log number of sequence states, at a given recognition density. At equilibrium ($\kappa = 0$), the potential (\ref{Psi_gamma_QG}) reduces to the well-known free fitness that generating the Boltzmann-Gibbs distribution of the substitution process (\ref{upm})~\cite{berg2004adaptive,sella2005application}. The ML likelihood entropy $S(\gamma^* (\ell), \ell)$ serves as a steepest-descent approximation for the Kulback-Leibler distance between the code density distribution $Q_{\gamma | \ell}$ and its neutral counterpart $Q^0_{\gamma | \ell}$, 
\EQA
H (Q_{\gamma | \ell} | Q^0_{\gamma | \ell}  ) & \simeq & S(\gamma^* (\ell), \ell) - S(\gamma_0, \ell)  
\nonumber \\
& \simeq  & \frac{\ell \, (\gamma^* (\ell) - \gamma_0)^2}{2 \gamma_0 (1 - \gamma_0)}, 
\EEA
which measures the information content of a regulatory motif of length $\ell$. By Sanov's theorem, the expected number of off-target sites in a random sequence of length $L$ is given by $L \exp[H (Q_{\gamma | \ell} | Q^0_{\gamma | \ell}  )]$, which sets a lower bound on the information content~\cite{gerland2002physical, lassig2007biophysics}. 

The extension to driven processes can be derived a weak-selection approximation, which is appropriate close to equilibrium. In this regime, we expand the rates $u_\pm (\gamma, \ell)$ and $\rho_\pm (\gamma)$ to first order in the selection coefficient $s_\gamma (\gamma, \ell)$. We can then write the stochastic dynamics of the mean coding density $\gamma$ as a diffusion process \cite{nourmohammad2013evolution,held2019survival}, 
\EQ
\dot \gamma = \frac{\Delta (\gamma, \ell)}{2N} \left ( (1 + \kappa) \frac{\partial S(\gamma, \ell)}{\partial \gamma} + 2N \frac{\partial F(\gamma, \ell)}{\partial \gamma} \right ) + \zeta (t). 
\label{gammadot} 
\EE
 The stochastic force $\zeta (t)$ has mean $\langle \zeta (t) \rangle = 0$ and covariance 
\EQ
\langle \zeta (t) \zeta (t') \rangle = \frac{\Delta (\gamma, \ell)}{2N}  \, \delta (t - t'), 
\EE
which is related to the diversity of the coding density in a population of size $N$, 
\EQA
\Delta (\gamma, \ell) & = & \frac{2N}{\ell} \, [\mu_+ (\gamma, \ell) + \mu_- (\gamma, \ell)] 
\nonumber \\
& = & \frac{2N \mu}{\ell} \left [ \gamma + \frac{\gamma_0}{1 - \gamma_0} (1 - \gamma) \right ]. 
\EEA
The neutral process includes recognition site and target mutations, leaving the neutral distribution $Q_0 (\gamma | \ell) = \exp[S(\gamma, \ell)]$ unaffected by the driving. In contrast, the population diversity $\Delta$ includes only site mutations.  This diversity governs the response to selection in Eq.~(\ref{gammadot}), while target mutations are independent of the fitness landscape $F(\gamma, \ell)$. Maximizing the potential $\Psi (\gamma | \ell)$ determines the conditional ML point $\gamma^* (\ell)$. This point satisfies a balance condition between selection, site mutations and the associated genetic drift, and target mutations. At equilibrium, this reduces to the well-known detailed balance between deleterious and beneficial substitutions (Fig.~2B). By Eqs.~(\ref{Psi_gamma_QG}) and (\ref{S}), the weak-driving balance condition relates selection and coding density, 
\EQ
2 N s^*_\gamma (\ell) = \frac{(1 + \kappa)}{\gamma_0 (1 - \gamma_0)} (\gamma^* (\ell) - \gamma_0). 
\label{sstar_traitW} 
\EE 
The evolutionary potential for length takes the weak-driving form 
\EQ
\Psi (\ell) \simeq \frac{\ep_0 \ell_0}{2} \, 2N \bar F_r (\ell) + 2N F_c (\ell) - \log Z_\ell, 
\label{Psi_ellW}
\EE
where $\ell_0 = \Delta G_{50} / (\ep_0 k_B T)$ and the recognition component, $\bar F_r (\ell)$, is determined by Eq.~(\ref{Fbar}). To derive this form, we use again a weak-selection  approximation and expand the substitution rates for extensions and compressions, $v_\pm (\gamma, \ell)$, to first order in the selection coefficients $s_\gamma$ and $c_0$. From Eqs.~(\ref{nupmpm})-- (\ref{vpm}), we obtain 
\EQA
v_+ (\gamma, \ell) & = & \nu ( 1 - Nc_0 ) + O(s_\gamma^2, s_\gamma c_0, c_0^2), 
\\
v_- (\gamma, \ell) & = &  \nu ( 1 - (\gamma - \gamma_0) \, N s_\gamma (\gamma, \ell) + Nc_0 ) 
\nonumber \\
& & + O(s_\gamma^2, s_\gamma c_0, c_0^2) .
\EEA
We can now compute the potential increment in the steepest-descent approximation for $\gamma$, 
\EQA
\lefteqn{\Psi (\ell) - \Psi (\ell - 1)}
\nonumber 
\\ & = & \log \frac{v_+ (\gamma^* (\ell - 1), \ell -1)}{v_- (\gamma^* (\ell), \ell)} 
\nonumber \\
& = & [\gamma^* (\ell) - \gamma_0] \, N s_\gamma^* (\ell) - 2N c_0 + O(s_\gamma^2, s_\gamma c_0, c_0^2)
\nonumber \\
& = & \frac{\ep_0 \ell_0}{2} \, [2N \bar F_r (\ell) - 2N \bar F_r (\ell - 1)] 
\nonumber \\
& & + 2N [F_c (\ell) - F_c (\ell - 1)]. 
\EEA
The last approximation uses that selection is proportional to the recognition load, $s_\gamma = \ep_0 (f_0 - F_r)$, and its ML value is approximately inversely proportional to the site length, $s_\gamma^* (\ell) \sim \ell_0 / \ell$, by Eqs.~(2) and~(\ref{sstar_traitW}). Eq.~(\ref{Psi_ellW}) then follows by summation of the increment up to a given value of $\ell$. Maximizing $\Psi (\ell)$ determines the balance condition for code length,
\EQ
\ell_0  \frac{\partial}{\partial \ell} s^*_\gamma(\ell) = -c_0.
\label{sstar_W}
\EE 
Together with Eq.~(\ref{sstar_traitW}) and the leading-order relation between coding density and length, Eq.~(2), we obtain an analytical closure for the global ML point $(\gamma^*, \ell^*)$ in the weak-driving regime, 
\EQ
\frac{\ell^*}{\ell_0} = \frac{1}{\gamma^* - \gamma_0} \simeq \sqrt{\frac{1 + \kappa}{\gamma_0 (1 - \gamma_0) \, 2N c_0}} . 
\EE
Compared to the effective fitness $\bar F (\ell)$, as given by Eq.~(\ref{Fbar}), the potential $\Psi (\ell)$ upweighs the recognition component, which implies that the global ML length $\ell^*$ overshoots the maximum of $\bar F (\ell)$ in the weak-driving regime (Fig.~4A). In Fig.~B1, we compare the full potentials $\Psi (\gamma | | \ell)$ and $\Psi (\ell)$ with their the weak-driving approximations; Fig.~B2 shows that this approximation produces the correct scaling of the equilibrium ML point with the cost parameter $c_0$. In particular, the condition that evolutionary equilibrium leads to near-maximal coding density, $\gamma^* (\kappa \! = \!0) \approx 1$, sets a constraint roughly equal to the selection for coding density, $2N c_0 \approx 2 N s^*_\gamma (\kappa \! = \!0) \approx 1$. This informs our choice of the cost parameter for the crossover to strong driving.\\

\subsection{Strong-driving regime} 
\begin{figure*}
  \begin{center}
  \includegraphics[scale=1]{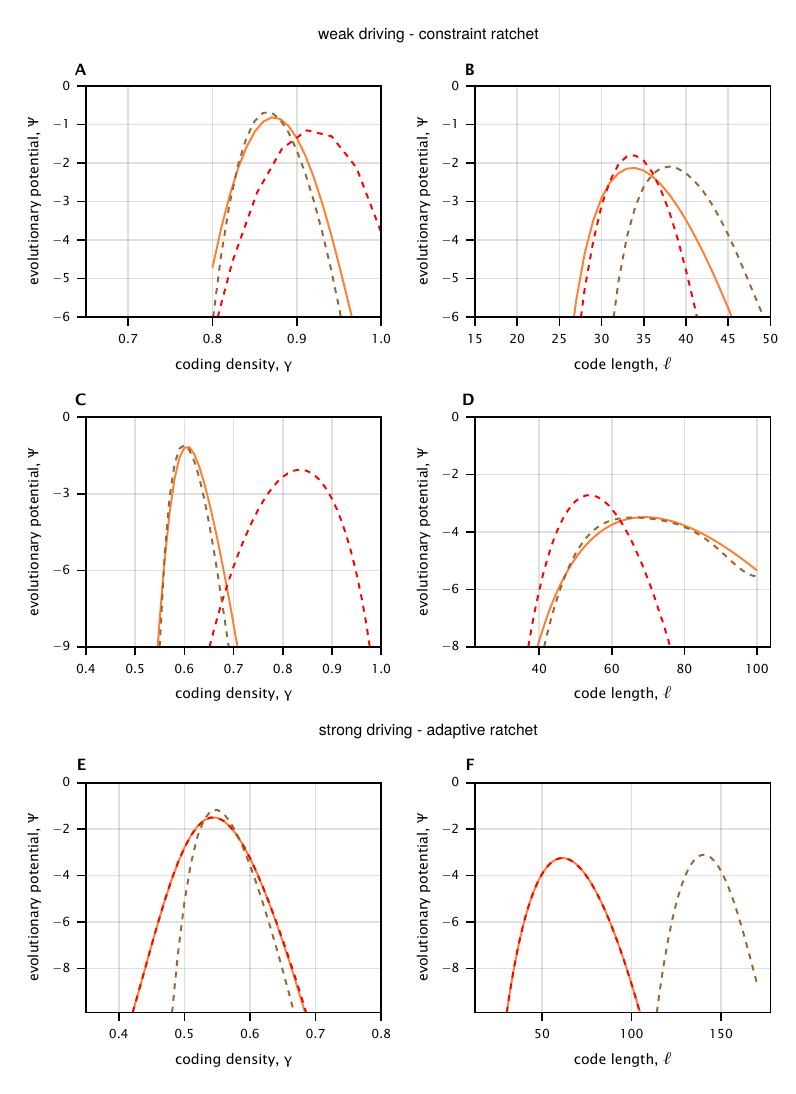}
  \end{center}
  
  \caption{
  {\small {\bf Approximate evolutionary potentials}. 
  (A),(C),(E) Evolutionary potential for coding density at ML code length, $\Psi (\gamma | \ell^*)$. 
  (B),(D),(F)Evolutionary potential for code length, $\Psi (\ell)$. Full potentials  (orange) are compared to the corresponding weak-driving (brown) and strong-driving (red) approximations. Parameters:
  (top)~weak-driving regime, $2N c_0 = 1, \kappa = 0.1$; 
  (center) weak-driving regime at low cost, $2N c_0 = 0.2, \kappa = 0.1$; 
  (bottom)~strong-driving regime, $2N c_0 = 1, \kappa = 15$.
  Other parameters as in Fig. 2.
  }
  }
  \label{fig:approx_potentials}
  \end{figure*}

\begin{figure*}[t]
    \begin{center}
    \includegraphics[scale=1]{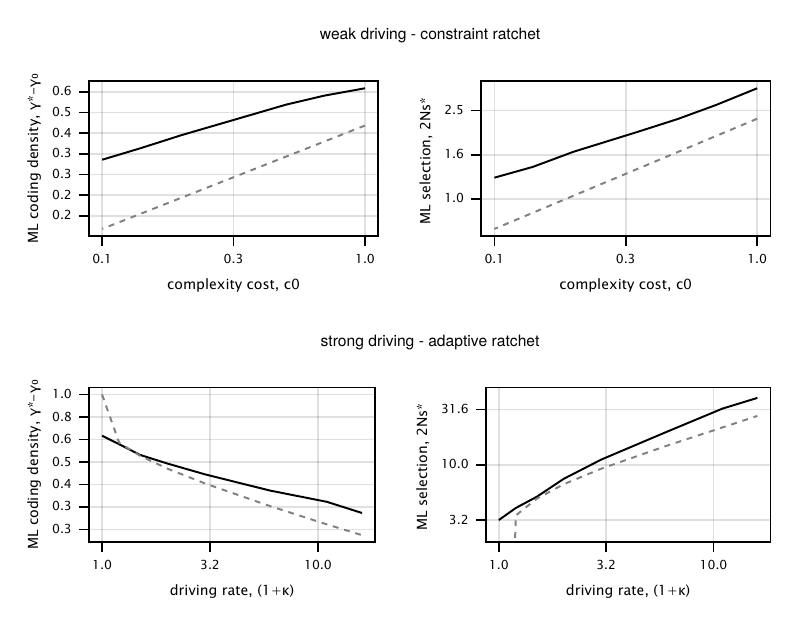}
    \end{center}
    
    \caption{
    {\small {\bf Global ML likelihood points.}
    (A, B)~Crossover to low cost. Full solution (solid) and weak-driving approximation (dashed) for coding density, $\gamma^* (2Nc_0, \kappa \! = \! 0)$ and selection on recognition, $s_\gamma^* (2N c_0, \kappa \! = \! 0)$. 
    (C, D)~Crossover to strong driving. Full solution (solid) and strong-driving approximation (dashed) for coding density, $\gamma^* (2Nc_0 \! = \! 1, \kappa)$ and selection on recognition, $s_\gamma^* (2N c_0 \! = \! 1, \kappa)$. 
    Other parameters as in Fig.~2. }
    }
    \label{fig:crossover}
    \end{figure*}

For $\kappa \gtrsim 1$, selection on code changes becomes strong ($s^*(\ell) \gg 1$), causing deleterious point mutations to freeze out. The resulting balance condition for the ML coding density involves beneficial site mutations and target mutations (Fig.~2D),
\EQ
    u_+(\gamma^*, \ell) + \rho_+(\gamma^*) = \rho_-(\gamma^*).
\EE
To first order,  we obtain
\EQ
\gamma_0 (1 - \gamma^* (\ell))(2 N s_\gamma^* (\ell) + \kappa ) =  \gamma^* (\ell)  (1 - \gamma_0) \, \kappa, .
\EE
Like its weak-selection counterpart, Eq.~(\ref{sstar_traitW}), this condition relates selection and coding density, 
\EQ
2N s_\gamma^* (\ell) = \frac{\gamma^*(\ell) - \gamma_0}{\gamma_0 (1 - \gamma^* (\ell))} \, \kappa . 
\label{sstar_trait1} 
\EE
Similarly, strong selection causes deleterious extensions and compressions to freeze out ($2N s_{+-} \ll 1$, $2 N s_{--} \ll 1$), and the balance condition for code length becomes 
\EQ
\gamma_0 s^*_{++}  =  (1 - \gamma^*) s^*_{-+}. 
\EE
Inserting the approximations (\ref{spmpm})  and solving for $s_\gamma^*$, we obtain
\EQ
s_\gamma^* = \frac{1 - (\gamma^* - \gamma_0)}{\gamma_0c_{++} - (1-\gamma^*)c_{-+}} \, c_0 , 
\label{sstar_length1} 
\EE
which is the strong-selection counterpart of Eq.~(\ref{sstar_W}).  The analytical closure for the global ML point in the strong-driving regime given by Eqs.~(\ref{sstar_trait1}) and~(\ref{sstar_length1}),
\EQA
0 & = & \frac{2Nc_0}{\kappa} \gamma_0 \left( 1 + \gamma_0 - \gamma^* \right)\left( 1-\gamma^*\right) 
\nonumber \\
& & + \left( \gamma^* - \gamma_0 \right) \left( c_{++} \gamma_0 - c_{-+}(1-\gamma^*) \right) , 
\EEA
determines the ML point $(\gamma^*, s^*)$; the ML code length, $\ell^*$, is then obtained from Eq.~(2). The full potentials $\Psi (\gamma | \ell)$ and $\Psi (\ell)$ and their strong-driving approximations are shown in Fig.~\ref{fig:approx_potentials}. The ML closure accurately captures the crossover to the strong-driving regime (Fig.~\ref{fig:crossover}). \\

\section{Dynamical analysis} 
\label{app_C}

\subsection{Ratchet evolution} 
\begin{figure*}
  \begin{center}
  \includegraphics[scale=1]{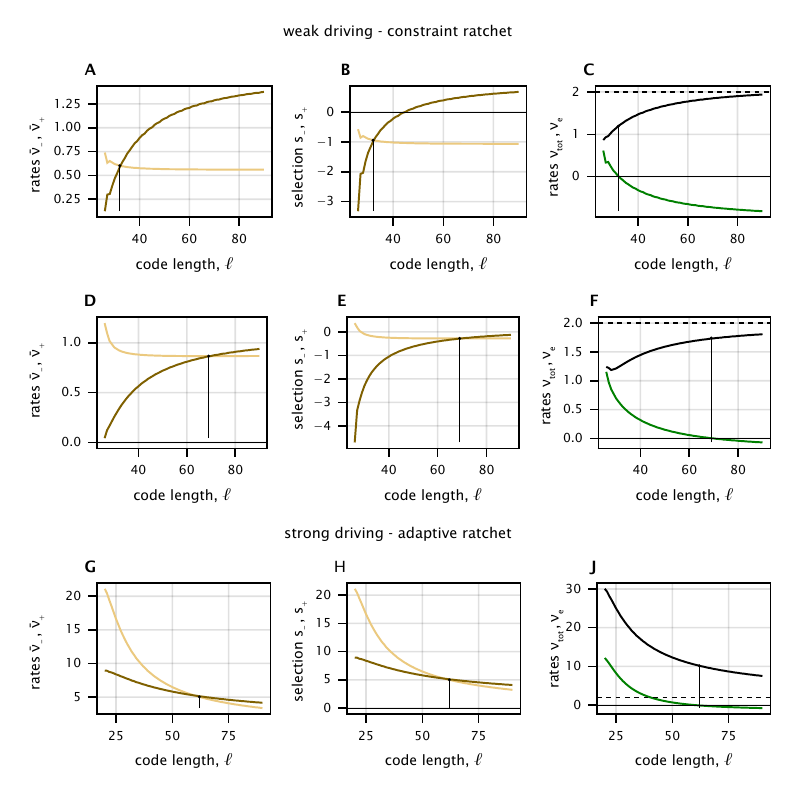}
  \end{center}
  
  \noindent 
  \caption{
  {\small {\bf Ratchet dynamics}. 
  (A)~Substitution rates for code extension and compression, $\bar v_+(\ell)$ (light) and $\bar v_- (\ell)$ (dark). 
  (B)~Ratchet selection coefficients for extension and compression, $s_+ (\ell)$ (light) and $s_- (\ell)$ (dark). 
  (C)~Total ratchet rate, $v_{\rm tot} (\ell)$ (black), and net elongation rate,  $v_e (\ell)$ (green). 
  Parameters: 
  (top)~weak-driving regime, $2N c_0 = 1, \kappa = 0.1$; 
  (center)~weak-driving regime at low cost, $2N c_0 = 0.2, \kappa = 0.1$; 
  (bottom)~strong-driving regime, $2N c_0 = 1, \kappa = 15$.
  Other parameters as in Fig. 2.}
  }
  \label{fig:ratchet_dyn}
  \end{figure*}

The ratchet dynamics derived in this paper is characterized by a breakdown of detailed balance for evolution under selection. Specifically, the minimal model for recognition sites generates net selection coefficients for code length changes, 
\EQ
2N s_\pm (\ell) = g^{-1} \left (\frac{\bar v_\pm (\ell)}{ \nu} \right ), 
\EE
where $\nu$ is the basic mutation rate of extensions and compressions, $\bar v_\pm (\ell)$ are the corresponding substitution rates defined by Eqs.~(\ref{vpm}) and~(\ref{Xbar}) (Fig.~\ref{fig:ratchet_dyn}). The function $g^{-1}$ is the inverse of the Haldane function $g$ defined in Eq.~(\ref{haldane}). Ratchet evolution of code length is defined by a decoupling of selection for reverse changes,
\EQ
s_+ (\ell) \neq s_- (\ell+1).
\EE
This inequality says that the substitution rates $\bar v_\pm (\ell)$ cannot be generated from the neutral rate $\nu$ and selection by any fitness landscape $F(\ell)$, which would imply   $s_+ (\ell) = - s_- (\ell + 1) = F(\ell + 1) - F(\ell)$ and a detailed balance relation of the form $v_+ (\ell) / v_- (\ell + 1) = g(s_+ (\ell)) / g(s_- (\ell + 1)) = \exp[2N (F(\ell + 1) - F(\ell)]$. In particular, the selection coefficients $s_\pm (\ell)$ decouple from the increment of the effective fitness landscape for code length, $\bar F (\ell + 1) - \bar F (\ell)$, as defined by Eq.~(\ref{Fbar}) (Fig.~4AC). This landscape maps the average long-term fitness effects of length changes but does not govern the ratchet dynamics. The selection coefficients $s_\pm (\ell)$ also decouple from the increment of the evolutionary potential, $\Psi (\ell + 1) - \Psi (\ell)$ (Fig.~4BD). In the stationary state of the ratchet, the substitution rates satisfy the detailed balance relation $\bar v_+ (\ell) / \bar v_- (\ell + 1) = \exp[\Psi (\ell + 1) - \Psi (\ell)]$, but the link of these rates to the neutral rate $\nu$ by Haldane's formula is lost.\\ 

\subsection{Adaptive tinkering} 
The ratchet substitution dynamics of extensions and compressions can be characterized by the total rate $v_{\rm tot}  (\ell) = \bar v_+ (\ell) + \bar v_- (\ell)$, which sets the overall speed of code length evolution, and the net elongation rate, $v_e (\ell) = \bar v_+ (\ell) - \bar v_- (\ell)$ (Fig.~\ref{fig:ratchet_dyn}). Inspection of the potential $\Psi (\ell)$, Eq.~(\ref{Psi_ell}) with rates given by Eq.~(\ref{vpm}), shows that the total rate takes the form $v^*_{\rm tot} (\ell) / (2 \nu) = 2N C(\ell) s^*_\gamma (\ell)$ with a slowly varying coefficient function $C( \ell) \sim 1$. In particular, we obtain an analytical estimate for the total ratchet rate at the ML point, as given by Eq.~(3). This relation implies  $v^*_{\rm tot} <  2 \nu$ in the weak-driving regime, where $2N s_\gamma \sim 1$, and $v_{\rm tot} > 2 \nu$ for $\kappa \gg 1$, in accordance with the evaluation from the full rates (Fig.~3CF, Fig.~\ref{fig:ratchet_dyn}). 

In the strong-driving regime, the range of adaptive tinkering around the ML point can be defined by the r.m.s.~code length fluctuations in the stationary state, 
\EQ
\Delta \ell = \Big ( \sum_\ell (\ell - \ell^*)^2 \, Q(\ell | \kappa) \Big )^{1/2} . 
\EE
We estimate $\Delta \ell$ from the curvature of the evolutionary potential for length, 
\EQA
\Delta \ell & = &  
\left (\frac{\partial^2 \Psi (\ell)}{\partial \ell^2} \, \Big |_{\ell^*}  \right )^{-1/2}
\nonumber \\ 
& \sim & \left (\frac{\partial \log [1 - \gamma^* (\ell)] }{\partial \ell} \, \Big |_{\ell^*} \right )^{-1/2}
\nonumber \\
& \sim & \ell^*, 
\label{broadness} 
\EEA
To derive this scaling, we Taylor-expand the potential $\Psi (\ell)$ around the ML point $\ell^*$, using Eq.~(\ref{Psi_ell}) with rates given by Eq.~(\ref{vpm}) in the strong-driving regime ($2N s_{+-}, 2 N s_{--} \ll 1$). Then we use the leading-order scaling $\gamma^* (\ell) - \gamma_0 \sim \ell^{-1}$, consistent with Eq.~(2). The analytical approximation of Eq.~(\ref{broadness}) is in accordance with the tinkering range evaluated from the full distribution $Q(\ell)$ (Fig.~3E). The ratchet generates anomalously broad length fluctuations compared to evolution in a uni-valued fitness peak of the form $F(\ell) = f_0 - c_r \ell^{- \alpha} - c_0 \ell$ ($\alpha > 0$). In that case, the stationary-state fluctuations scale differently, $\Delta \ell \sim (\ell^*)^{1/2}$, as can be seen by evaluating the Boltzmann-Gibbs distribution $Q_{\rm eq} (\ell) \sim \exp[2N F(\ell)]$. \\

\begin{figure*}
  \begin{center}
  \includegraphics[scale=1]{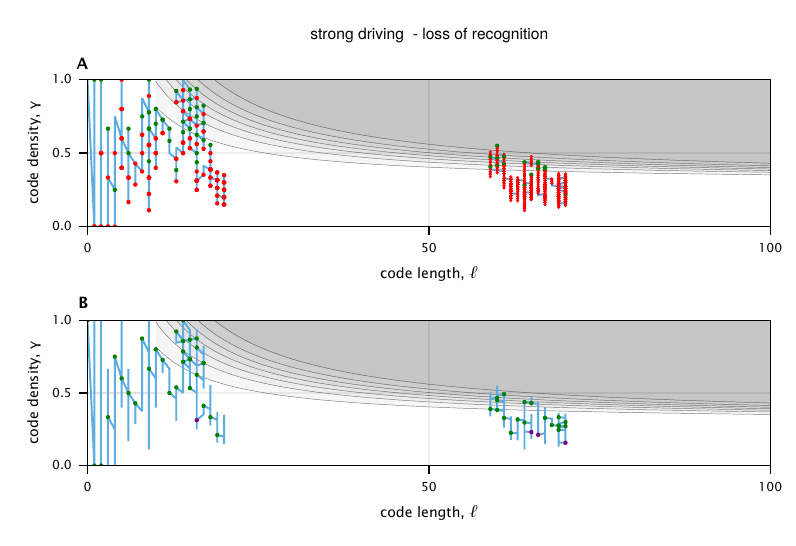}
  \end{center}
  
  \caption{
  {\small {\bf Evolutionary paths with loss of function}. 
  Loss of function by degradation at low code length (left, $2Nc_0 = 10$, $\kappa = 15$, 2000 mutational time steps, starting at $\ell = 20$) and at 
  high code length (right, $2N c_0 = 1$, $\kappa = 100$, 2000 mutational time steps, starting at $\ell = 70$). 
  (A) Coding density changes (green: adaptive recognition site mutations, red: recognition target mutations).
  (B) Code length changes (green: adaptive, purple: deleterious). 
  Other parameters as in Fig.~2. }
  }
  \label{fig:loss_traj}
  \end{figure*}

We note that the state of adaptive tinkering at high molecular complexity is transient because it is subject to an instability: with a small probability, genetic drift can cause loss of function ($\gamma < \gamma_{50} (\ell)$); subsequently, the complexity cost $F_c (\ell)$ generates rapid decay of dysfunctional sites. Fig.~\ref{fig:loss_traj} shows trajectories ending with loss of function occurring at different length scales.

\subsection{Number of adaptive paths and fitness flux} 
The effective number of adaptive paths for a given quantitative trait, here $\Delta G$, is defined as its suitably normalized heritable variation, 
\EQ
\A (\gamma, \ell) = \frac{{\rm Var} (\Delta G / k_B T)}{2 \mu N \ep_0^2}, 
\EE
where $2 \mu N$ is the neutral sequence diversity per unit of length and $\ep_0^2 = {\rm Var} (\Delta \Delta G / k_BT)$ is the variance of the mutational effect distribution. 
The number of adaptive paths is a measure of susceptibility to selection \cite{lynch1998genetics}. This can be seen by relating it to the change in mean population fitness by natural selection, $\phi_s \equiv \left. \mbox{$\frac{d}{dt}$} \right |_{\rm sel} E (F)$, in a trait-dependent fitness landscape $F(\Delta G)$. By Fisher's theorem, we obtain 
\EQA
\hspace*{-05cm}  \phi_s & = & {\rm Var }(F_r) 
\nonumber \\
& = & 2 \mu N \ep_0^2 (F'_r (\Gamma)^2 \,  \A(\gamma, \ell) + O(F''_r (\Gamma) \, {\rm Var }F_r), 
\label{Fisher} 
\EEA
where $\Gamma = E(\Delta G / k_BT)$ denotes the population mean trait. The selective flux $\phi_s $ is a non-negative component of the fitness flux $\phi$. In an evolutionary steady state,  $\phi_s $ measures the amount of selective turnover required to maintain a time-independent average trait $\langle \Gamma \rangle$. The total fitness flux $\phi$ also includes the fitness change by evolution under mutations, selection, and genetic drift \cite{mustonen2010fitness}. It vanishes in an evolutionary equilibrium state. In a driven stationary state, it has a time-independent average, $\langle \phi \rangle > 0$, that balances with the rate of fitness change by external fluctuations. 

In a model with site-dependent trait effects $\ep_i$ ($i = 1, \dots, \ell$), we obtain an effective number of adaptive paths
\EQ
\A
\simeq \left \{
\begin{array}{ll} 
\sum_{i=1}^\ell \frac{\ep_i^2}{\ep_0^2}  \left ( \frac{\gamma_0}{1 - \gamma_0} (1 - \gamma_i) + \gamma_i \right ) & \mbox{(weak selection)}
\\ \\
\sum_{i=1}^\ell \frac{\ep_i^2}{\ep_0^2}  \left ( \frac{\gamma_0}{1 - \gamma_0} (1 - \gamma_i) \right ) & \mbox{(strong selection)} , 
\end{array} \right . 
\EE
where $\gamma_i = 1$ for site positions with a target match and $\gamma_i = 0$ for positions with a mismatch. The weighting factors $\ep_i^2/\ep_0^2$ reflect the higher contributions of strong-effect mutations to the adaptive process; different regimes for transient adaptive dynamics have been discussed in ref.~\cite{Kopp07}. In the weak-selection regime ($2 N \ep_0 F'(\Gamma) \lesssim 1$), all positions contribute an expected sequence diversity $2N \mu$, generating the maximum trait diversity \cite{lynch1998genetics,nourmohammad2013evolution}. In the strong-driving regime ($2 N \ep_0 F'(\Gamma) \gg 1$), the trait diversity narrows to positions with deleterious majority alleles ($\gamma_i = 0$). These positions  contribute an expected trait diversity $2 \mu N \frac{\gamma_0}{1 - \gamma_0} \ep_i^2$, predominantly from mutations destined for fixation.  Positions with beneficial majority alleles ($\gamma_i =1$) freeze out; these positions have a much smaller trait diversity $\mu \ep_i^2 /(\ep_i F'(\Gamma))$. Together, the number of adaptive paths $\A$ depends on the distribution of effects $\ep_i$ and the occupancy of beneficial and deleterious sequence states but decouples from the global parameters $\mu$, $N$, and from the overall strength of selection, $F' (\Gamma)$. 

Here we focus on the expected number of adaptive paths in an ensemble of parallel-evolving populations, $A = \langle \A \rangle$. In the strong-selection regime of the minimal model ($\ep_i = \ep_0, \langle \gamma_i \rangle = \gamma$), we obtain
\EQ
A(\gamma, \ell)  = \frac{\gamma_0}{1 - \gamma_0} (1 - \gamma) \,  \ell .
\label{A_SC}
\EE
At the global ML point $(\gamma^*, \ell^*)(\kappa)$, this relation reduces to Eq.~(4); the function $A^* (\kappa)$ is shown in Fig.~4F. The number of adaptive paths can be compared to the expected adaptive speed of the recognition trait by point substitutions,
\EQ
\langle \phi_{G} \rangle (\gamma, \ell) = \ell \ep_0 \, [u_+ (\gamma, \ell) - u_- (\gamma, \ell)] ,
\EE
and the corresponding fitness flux,  
\EQ
\langle \phi \rangle (\gamma, \ell)  
= \ell \, [u_+ (\gamma, \ell) - u_- (\gamma, \ell)]  \, s_\gamma (\gamma, \ell); 
\EE
the contribution of length changes to the adaptive fluxes is down by a factor $O(\nu/\mu)$. 
In the strong-driving regime, we obtain 
\EQA
\langle \phi_G \rangle (\gamma, \ell) & \simeq &   \ell \ep_0  \, u_+(\gamma, \ell) 
\nonumber \\
& = & 2 \mu N \ep_0 s_\gamma (\gamma, \ell) \, A (\gamma, \ell)
\label{phi_ave_G}
\EEA
and 
\EQA
\langle \phi \rangle (\gamma, \ell) & \simeq &   \ell \, u_+(\gamma, \ell) \, s_\gamma (\gamma, \ell)
\nonumber \\
& = & 2 \mu N  s_\gamma^2 (\gamma, \ell) \, A (\gamma, \ell)
\label{phi_ave}
\EEA
using Eq.~(\ref{upm}). Thus, at a given value of $s_\gamma$, the expected fluxes $\langle \phi_G \rangle$ and $\langle \phi \rangle$ increase with $A$, while the selection $s_\gamma$ required for a given value of $\langle \phi_G \rangle$ decreases with increasing $A$. In an evolutionary steady state, the adaptive fluxes match the effect of recognition target changes, as given by the  strong-driving balance condition (\ref{sstar_trait1}). We obtain 
\EQA
\langle \phi_G^* \rangle (\ell)  & = & \mu \kappa  \ell \ep_0 \, (\gamma^*(\ell) - \gamma_0)
\nonumber \\
& \simeq &\mu \kappa  \ell_0 \ep_0 ,
\EEA
and 
\EQA
\langle \phi^* \rangle (\ell)  & = &\mu \kappa \ell \, (\gamma^*(\ell) - \gamma_0) \, s_\gamma^* (\ell)
\nonumber \\
& = & \frac{\mu \kappa^2 \ell}{2N} \frac{ (\gamma^*(\ell) - \gamma_0)^2}{\gamma_0 (1 - \gamma^*(\ell))} 
\nonumber \\
& \simeq & \frac{\mu \kappa^2}{2N} \frac{\ell_0^2}{(1 - \gamma_0)} \frac{1}{ A^*(\ell)}, 
\EEA
where we use the leading-order scaling $\gamma^* (\ell) - \gamma_0 \sim \ell_0/\ell$, consistent with Eq.~(2). According to these relations, the adaptive speed of the recognition trait becomes asymptotically independent of $\ell$; the fitness flux required to maintain the stationary state decreases with increasing number of adaptive paths. \\

\section{Immune recognition} 
 \label{app_D}
 
\subsection*{Recognition complexity} 

The effective number of epitopes targeted by an adaptive immune response, $g_r$, depends on the total number of accessible epitopes, $g$, presented by the pathogen and on stochastic differences in immunogenicity, which are characteristics of the host's immune repertoire. For simplicity, we assume the pathogen has $g$ independent, equally accessible, and statistically equivalent epitopes. In this case, the potencies $\Z = (Z_1, \dots, Z_g)$ of a given host are independent random variables drawn from a common distribution $Q(Z)$, which determines the population statistics of adaptive immune response to a given pathogen. Specifically, we compute the population-averaged log recognition complexity, $\log g_r \equiv \langle S(\z) \rangle$, where $S(\z)$ denotes the Shannon entropy of the normalized potency distribution, $S(\z) = - \sum_{i=1}^g z_i \log z_i$ with $z_i = Z_i / \sum_{i=1}^g Z_i$. 

A key feature of adaptive immune response is the density of B cell lineages with genetically different receptors, $\Omega (K)$, as a function of the binding constant $K$ to a given epitope. The extremal-value statistics of this density of states determines two important repertoire parameters~\cite{morantovar2024primaryresponse}: (i) the expectation value of strongest binding in a repertoire of $L_0$ lineages, $K^* = \langle \min K \rangle$, and the repertoire exponent $\beta^* \equiv (K d/dK) \log \Omega (K) |_{K^*}$. This exponent is approximately determined by the repertoire size $L_0$, the code length of the binding locus, $\ell$, and the affinity variance in an ensemble of random sequences, $\sigma^2_{\log K, 0} \equiv \ell \ep_0^2$, where $\ep_0$ denotes the affinity effect scale of point mutations in the receptor sequence. We obtain~\cite{morantovar2024primaryresponse}  
\EQ 
\beta^* \simeq \frac{\sqrt{2\log L_0}}{\ep_0 \ell^{1/2}}. 
\label{eq:beta*}
\EE
At constant binding parameters $\ell$ and $\ep_0$, the repertoire exponent $\beta^*$ increases with increasing repertoire size $L_0$. At constant $L_0$, the exponent decreases with increasing $\ell$ or $\ep_0$. Hence, by evolutionary changes of $L_0$, $\ell$, and $\ep_0$, host systems can independently vary the repertoire parameters $K^*$ and $\beta^*$.

The repertoire exponent $\beta^*$ determines the fluctuations of epitope-specific effective affinities $\log Z \equiv \Delta G/(k_bT)$, as given by the variance $\sigma^2_{\log Z}$. To compute this variance, we use three results of~ref.~\cite{morantovar2024primaryresponse}: (i)~Fluctuations in the maximum affinity have variance $\sigma^2_{\log K^*} = \pi^2/(6 \beta^{*2})$. (ii)~In an exponential infection process, affinity changes induce correlated changes in B cell clone size, $N \sim (K^*)^{-\zeta/\beta^*}$ with $\zeta/\beta^* \approx 1$ for physiological parameters of human repertoires. (iii)~Antiserum potencies are given by $Z \sim N/K^*$. Together, we obtain 
\EQ
\sigma^2_{\log Z}  = \sigma^2_{\log (N/K^*)} \approx \frac{4 \pi^2}{6}\frac{1}{\beta^{*2}}. 
\label{eq:VarlogZ}
\EE

\begin{figure}[t]
  \begin{center}
  \includegraphics[width=0.9\linewidth]{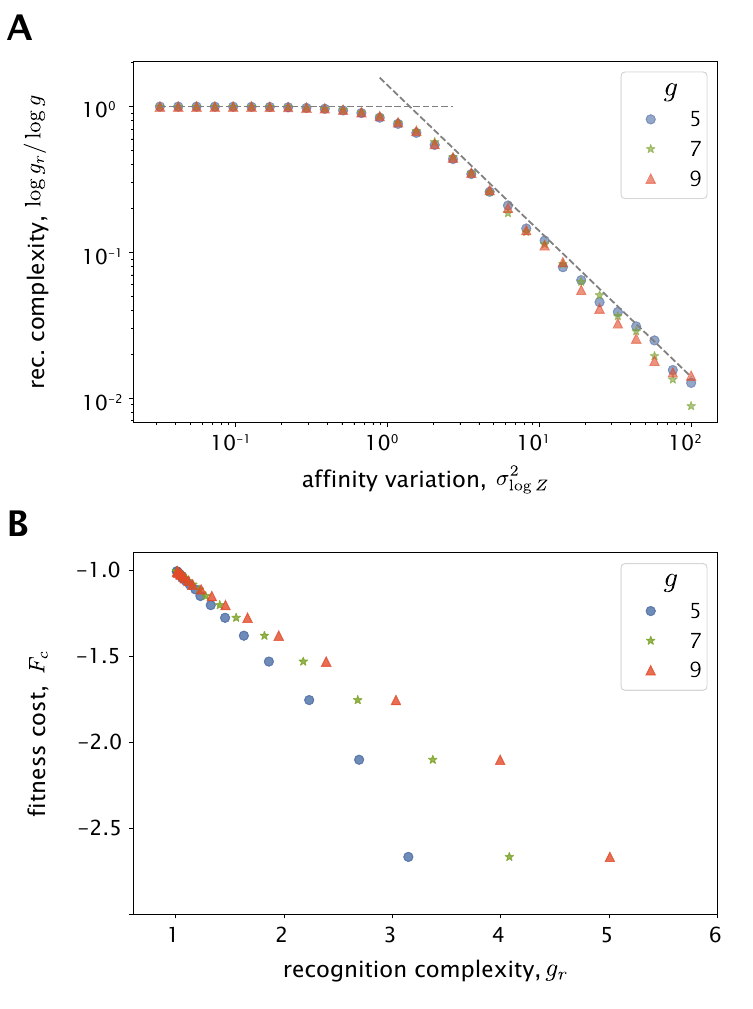}
\end{center}
\caption{
  {\small {\bf Repertoire dependence and fitness cost of recognition complexity. }
  (A)~Scaled log recognition complexity, $\log g_r / \log g = \langle S(\z) \rangle$, plotted as a function of the target affinity variation $\sigma^2_{\log Z}$ together with the asymptotic forms given by Eq.~(\ref{g_r}) (dashed lines). The population-averaged entropy is obtained numerically by sampling 1000 potency distributions $\z$ from the immune response distribution described in the text~\cite{morantovar2024primaryresponse}. 
 (B)~Host fitness cost, $F_c (g_r)$. This function is derived from the fitness cost of repertoire size $L_0$, as explained in the text. Repertoire parameters: $\beta^* = 1/\zeta = 2.5$; see refs.~\cite{bocharov1994bcells, morantovar2024primaryresponse, smith2010viral, pawelek2012viral, goyal2021viral, sender2021viral}. 
  }}
  \label{fig:g_r}
  \end{figure}

Next, we evaluate the recognition complexity by numerical simulations of stochastic lineage activation~\cite{morantovar2024primaryresponse}, generating an ensemble of potency distributions $\z$ in a repertoire with parameters $L_0$, $\ell$, and $\ep_0$. We obtain a remarkably simple scaling form, 
\EQ
\log g_r \simeq  (\log g) \times 
\left \{ 
\begin{array}{ll}
1 & (\sigma^2_{\log Z} \lesssim 1)
\\ \\
 (\sigma^2_{\log Z})^{-1}  & (\sigma^2_{\log Z} \gtrsim 1), 
 \end{array} \right. 
 \label{g_r} 
\EE
as shown in Fig.~\ref{fig:g_r}A. This form describes a cross-over from a high-complexity, low-immunodominance regime ($g_r \approx g$ for $\sigma^2_{\log Z} \lesssim 1$) to a low-complexity, high-immunodominance regime set by immunogenicity fluctuations of the host repertoire ($g_r < g$ for $\sigma^2_{\log Z} \gtrsim 1$). 

The preceding analysis also establishes a generic fitness cost of recognition complexity, defining a cost component $F_c (g_r)$ in the fitness model of Eq.~(\ref{F}). Assuming a fitness cost of repertoire size $L_0$, we can compute the resulting fitness landscape $F_c (g_r)$ by combining the repertoire relations~(\ref{eq:beta*}) -- (\ref{g_r}). For example, a repertoire fitness cost of order $\log L_0$ (e.g., induced by off-target binding) produces the landscape $F_c (g_r)$ shown in Fig.~\ref{fig:g_r}B, which describes a monotonically increasing, approximately linear cost over the relevant range of $g_r$ (cf.~Fig.~6). Other assumptions produce qualitatively similar landscapes, the details of which are beyond the scope of this paper. In the minimal fitness model discussed used in the main text, we use the linear approximation $F_c (g_r) = c_0 g_r$. 
 
\subsection*{Recognition dynamics} 

Our minimal dynamics model has three characteristics: 
(i)~Functional binding of antibodies to an antigenic epitope leads to neutralization of the antigen. The neutralization probability $R$ is described by a Hill recognition function depending on the reduced binding affinity $\Delta G / (k_BT)$, as given by Eq.~(\ref{Hill})~\cite{hobson1972titersprotection, coudeville2010titersprotection}. 
(ii)~An infection at time $t_0$ targets $a$ epitopes with initially equal affinities,  $\Delta G_\alpha (t_0) / (k_BT) = \Delta G_i / (k_BT) = \Delta G_{50} / (k_BT) +  2\ep_0$ ($\alpha = 1, \dots, g$).  This sets the entropy $S(z) = \log g$ and the recognition complexity $g_r = g$. 
(iii)~Antigenic drift is a Poisson process that changes epitope sequences with a point substitution rate $\rho$; each epitope mutation reduces the affinity by a constant amount, $\ep_0 = \Delta \Delta G / k_BT$. Thus, antigenic drift generates time-dependent affinities $\Delta G_\alpha (t)$ ($\alpha = 1, \dots, g$). 
(iv)~Exposure of an individual to the evolving antigen is an independent Poisson process with rate $\tau^{-1}$. For an exposure event at time $t$, we evaluate the recognition function
\EQ
R (t) = 1 - \prod_{\alpha = 1}^g \left(1-\frac{1}{ 1+\exp \left [\frac{1}{k_BT} (\Delta G_{50} - \Delta G_\alpha (t)) \right ]}\right). 
\label{Rt} 
\EE
Neutralization by activation of immune memory occurs with probability $R(t)$, given that each epitope is an independent and sufficient channel for recognition of the antigen~\cite{koel2013singlemutation}. A full secondary infection occurs with probability $\Delta R (t) = 1 - R(t)$, resetting all affinities to $\Delta G_0$. Here we assume that significant updating of the immune response repertoire occurs only after full infections. Additionally, exposure events that occur between full infection events can boost serum antibody titers against the corresponding pathogen~\cite{zarnitsyna2016multiepitope}. However, this does not affect our results, which rely on memory response rather than on serum antibodies.

We simulate long-term affinity trajectories $(\Delta G_1, \dots, \Delta G_g) (t)$  for the joint dynamics of antigenic drift and exposures. The effective speed of antigenic evolution, $\kappa \equiv \rho \tau$, is given by the expected number of mutations per epitope between subsequent exposures. We record the steady-state average antigenic fitness component $F_r (g, \kappa) = f_0 \langle \Delta R \rangle (g, \kappa)$; this quantity is proportional to the fitness cost of infections. Together with a complexity cost of the form $F_c (g) = - c_0 g$, we obtain a minimal host fitness function of the form~(2).

The conditional fitness landscape $F(g | \kappa)$ determines the optimal epitope complexity for given antigenic speed, $g^*(\kappa)$. For low to moderate antigenic drift ($\kappa \lesssim 2$), $g^* (\kappa)$ increases with $\kappa$ (Fig.~5B). For higher values of $\kappa$, the number of epitopes needed for protection increases exponentially, inducing a prohibitive cost and a transition to $g^* (\kappa) = 0$. A negative correlation between the speed of antigenic evolution and the efficacy immune memory has also been mapped in multi-pathogen models~\cite{Mayer2019}.

\subsection*{Code Availability}
All code used for numeric computations and to generate figures can be found in a Github repository (\url{https://github.com/tomroesch/complexity_evolution/tree/Publication}).


\twocolumngrid
\bibliography{./complexity_references.bib}

\end{document}